\documentclass{article}
\usepackage[title]{appendix}
\usepackage{graphicx}
\usepackage{subcaption}
\usepackage{multicol}
\usepackage{xcolor}
\usepackage{verbatim}
\usepackage{amsmath,amssymb,amsthm}
\usepackage{arxiv}
\usepackage{hyperref}
\usepackage{multirow}
\usepackage{algorithm}
\usepackage[noend]{algpseudocode}
\usepackage{arydshln}
\usepackage{booktabs}
\newtheoremstyle{thm}
{9pt}
{9pt}
{\itshape}
{}
{\bfseries}
{.}
{ }
{}

\theoremstyle{thm}

\newtheorem{theorem}{Theorem}[section]

\newtheorem{corollary}[theorem]{Corollary}

\newtheorem{remark}[theorem]{Remark}

\newenvironment{prf}{\textbf{\emph{Proof.}}}{\qed}

\newcommand{\R}{\mathbb{R}}

\newcommand{\N}{\mathbb{N}}
\newcommand{\E}{\mathbb{E}}
\newcommand{\PP}{\mathbb{P}}

\title{Change-point analysis of textual data: a flexible approach}

\author{
 \href{https://orcid.org/0000-0001-8243-9794}
 {\includegraphics[scale=0.06]{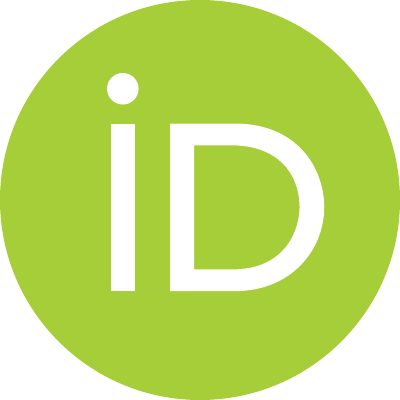}\hspace{1mm}Bojana Milo\v sevi\'c}\\
 University of Belgrade\\
 Faculty of Mathematics\\
 \texttt{bojana@matf.bg.ac.rs}
 \And
 \href{https://orcid.org/0009-0001-7275-6654}
 {\includegraphics[scale=0.06]{orcid.pdf}\hspace{1mm}Luka Perovi\'c}\\
 University of Belgrade\\
 Faculty of Mathematics\\
 \texttt{luka.perovic@matf.bg.ac.rs}
}

\begin{document}

\maketitle

\begin{abstract}
We propose a CUSUM-based change-point procedure for word-frequency vectors that accounts for the variation in the amount of observed textual information over time. Under a multinomial model, we derive the asymptotic null distribution of the proposed statistic, allowing for varying document numbers and lengths. With trimming based on cumulative word counts, the limiting null distribution does not depend on the temporal distribution of textual information.
We also consider parametric and nonparametric bootstrap procedures for finite-sample calibration. A simulation study examines their performance under both multinomial sampling and Dirichlet mixtures of multinomial distributions. The results demonstrate the advantages of the proposed approach under increasing document intensity and show that resampling entire documents provides more reliable calibration in the presence of document-level heterogeneity. Finally, we apply the procedure, combined with binary segmentation, to 68,542 abstracts from the arXiv category stat.ML spanning 2007–2025, identifying multiple changes in word-usage patterns.

\keywords{multivariate count data; CUSUM statistics;word frequencies; stat.ML}
\end{abstract}

\section{Introduction}
\sloppy

Change-point analysis provides a fundamental framework for detecting
structural changes in data-generating mechanisms. In the context of textual
data, this problem is particularly challenging because text is not inherently
numerical and must first be transformed into an appropriate statistical
representation.

Among the most common approaches, documents are represented through
bag-of-words frequency vectors and modeled using multinomial distributions.
In this framework, change-point detection aims to identify temporal changes
in word-usage patterns. Such models are attractive because of their
simplicity, interpretability, and direct connection to document frequencies (see \cite{salton1975vector}, \cite{preis2024change}).

Several approaches to change-point detection in textual data have been developed by exploiting either direct word-count representations or lower-dimensional latent topic representations. In count-based approaches, documents are represented by vectors of word frequencies and modeled as multinomial observations, so that a structural change corresponds to a change in the underlying vector of word probabilities. General change-point procedures for multinomial sequences have been studied, for example, by  \cite{horvath1995testing},\cite{batsidis2013change}. More recently,\cite{preis2024change} proposed a change-point procedure specifically tailored to text documents represented by absolute word-frequency vectors and modeled using multinomial distributions.

An alternative line of work is based on topic models, where each document is represented through a distribution over a smaller number of latent topics rather than directly through the full vocabulary. Changes in a document stream can then be interpreted as changes in these topic proportions. \cite{wang2018realtime}, for instance, combined online latent Dirichlet allocation with Bayesian online change-point detection to identify real-time changes in streams of news articles and social-media documents. \cite{bose2021changepoint}considered the offline setting and developed a specialized temporal topic model in which change-points correspond directly to changes in the distribution of topic proportions. Their procedure first estimates the underlying topic structure and subsequently employs likelihood-ratio statistics together with a modified wild binary segmentation algorithm to locate multiple change-points. More recent work has moved towards integrating topic evolution and change-point detection within a single model; for example, \cite{wang2025fine} proposed a Pitman-Yor-process-based framework for detecting fine-grained, topic-specific changes over time. 

{Beyond change-point detection, statistical inference
for textual data has also been considered without
relying on a fully specified generative model.
For example, \cite{PhelanCampbell2026} combined
nonnegative matrix factorization with regression
and bootstrap-based inference to investigate
the effects of document-level covariates on
topic prevalence.}

While multinomial change-point models provide a natural framework for
textual data analysis, their practical applicability is often limited by two
important factors. First, real-world document collections rarely follow an
exact multinomial model and frequently exhibit additional variability.
Second, the number of available documents often varies substantially over
time and may even display a strong increasing trend, particularly in
bibliometric applications (see \cite{gusenbauer2021age}). As we show, such heterogeneity can considerably affect the
performance and calibration of existing procedures.

Motivated by these challenges, we propose cumulative-sum change-point statistics for word-frequency vectors that explicitly account for variation in the amount of textual information observed over time. We derive their asymptotic null behavior under the multinomial model and introduce bootstrap calibration procedures suitable for settings with varying numbers and lengths of documents. We further investigate robustness under Dirichlet mixtures of multinomial distributions, allowing for additional document-level heterogeneity beyond the multinomial model.

The remainder of the paper is organized as follows. Section \ref{sec2} introduces the proposed change-point statistics, while Section \ref{sec3} establishes their asymptotic properties. Section \ref{sec4} evaluates their finite-sample performance through an extensive simulation study and compares them with existing procedures. Finally, Section \ref{sec5} illustrates the methodology using arXiv abstracts from 2007 to 2025.

\section{Proposed methodology}
\label{sec2}

We assume that documents, which are represented by their word frequency vectors, arrive at different points in time in an observed time period. For our analysis, we group these documents into a discrete number of groups, each corresponding to a smaller time frame in the observed time period. In further notation, we treat these time frames as months. For our theoretical analysis, this can be considered completely arbitrary and is only left in for the purpose of interpretability.

Before presenting the proposed statistics, we introduce the following
notation:
\begin{itemize}
    \item $V$ is the number of selected words that are followed through time;
    \item $X_{ij}\in\N_0^V$ is the word-frequency vector of $V$ selected words associated with the
    $j$th document in month $i$;
    \item $\pi_{ij} \in \mathbb{R}^V$ is the assumed vector of the multinomial probabilities of each selected word for the $j$th document in month $i$;
    \item $N_i$ is the number of documents observed in month $i$, with
    $\sum_{i=1}^T N_i=N$;
    \item $n_{ij}$ is the total number of words in the $j$th document from
    month $i$;
    \item $T$ is the number of observed time periods (months);
    \item
    \[
    m_i=\sum_{j=1}^{N_i}n_{ij},
    \qquad
    M_k=\sum_{i=1}^k m_i,
    \qquad
    M_T=\sum_{i=1}^T m_i
    \]
    are, respectively, the number of words in month $i$, the cumulative
    number of words up to month $k$, and the total number of words.
\end{itemize}

Under the null hypothesis, we assume that no change occurs in the distribution of the selected words over time, that is:
$H_0: \pi_{ij} = \pi$ for each $i \in \{1,..,T\}$ and $j \in\{1,...,N_i\}$, where $\pi$ is unspecified. The alternative hypothesis is complementary to the null.

Let
\[
C_k=\sum_{i=1}^k\sum_{j=1}^{N_i}X_{ij},
\qquad k=1,\ldots,T,
\]
be the cumulative frequency vector of the selected words. Under the null
hypothesis of no change, the vector of probabilities of appearance of the
selected words is estimated by
\begin{align}\label{ocenaPi}
\widehat{\pi}
=
\frac{C_T}{M_T}
=
\frac{1}{\sum_{i=1}^T\sum_{j=1}^{N_i}n_{ij}}
\sum_{i=1}^T\sum_{j=1}^{N_i}X_{ij}.
\end{align}
The corresponding cumulative-sum vector is
\begin{equation}\label{eq:dk}
d_k
=
C_k-M_k\widehat{\pi}
=
C_k-\frac{M_k}{M_T}C_T.
\end{equation}

We propose the statistic
\begin{equation}\label{statB}
B
=
\max_{\frac{k}{T}\in[\varepsilon,1-\varepsilon]}
\frac{d_k^\top\widehat{\Sigma}_p^{-1}d_k}
{\dfrac{M_k(M_T-M_k)}{M_T}},
\end{equation}
where
\[
\widehat{\Sigma}_p
=
\operatorname{diag}(\widehat{\pi}_1,\ldots,\widehat{\pi}_V)
-\widehat{\pi}\widehat{\pi}^{\top}.
\]
Writing
\[
\widehat{\pi}_{V+1}
=
1-\sum_{r=1}^V\widehat{\pi}_r
\]
for the estimated probability of all other corpus words, we have
\begin{equation}\label{eq:inverse-sigma-p}
\widehat{\Sigma}_p^{-1}
=
\operatorname{diag}\left(
\frac{1}{\widehat{\pi}_1},\ldots,
\frac{1}{\widehat{\pi}_V}
\right)
+
\frac{1}{\widehat{\pi}_{V+1}}
\mathbf{1}_V\mathbf{1}_V^\top,
\end{equation}
provided that all $\widehat{\pi}_1,\ldots,\widehat{\pi}_{V+1}$ are
strictly positive. Equivalently,
\[
\widehat{\Sigma}_p^{-1}
=
\begin{bmatrix}
\dfrac{1}{\widehat{\pi}_1}
+\dfrac{1}{\widehat{\pi}_{V+1}}
&
\cdots
&
\dfrac{1}{\widehat{\pi}_{V+1}}
\\[2mm]
\vdots
&
\ddots
&
\vdots
\\[2mm]
\dfrac{1}{\widehat{\pi}_{V+1}}
&
\cdots
&
\dfrac{1}{\widehat{\pi}_V}
+\dfrac{1}{\widehat{\pi}_{V+1}}
\end{bmatrix}.
\]

\section{Asymptotic results}
\label{sec3}
We first establish the asymptotic null distribution of the cumulative-sum
process underlying statistic $B$.

For each occurrence of a word in the corpus, define a random vector
$Z_\ell\in\R^V$ by
\[
Z_\ell
=
\begin{cases}
e_r, & \text{if the $\ell$th word is the $r$th selected word},
       \quad r=1,\ldots,V,\\
0,   & \text{if the $\ell$th word is any other corpus word},
\end{cases}
\]
where $e_r$ denotes the $r$th canonical basis vector in $\R^V$.

Under the null hypothesis, assume that $Z_1,Z_2,\ldots$ are independent and
identically distributed with
\[
\PP(Z_\ell=e_r)=\pi_r,
\qquad r=1,\ldots,V,
\]
and
\[
\PP(Z_\ell=0)=\pi_{V+1}
=
1-\sum_{r=1}^V\pi_r.
\]
Then
\[
\E Z_\ell=\pi=(\pi_1,\ldots,\pi_V)^\top
\]
and
\begin{equation}\label{eq:sigma-p}
\Sigma_p
=
\operatorname{Var}(Z_\ell)
=
\operatorname{diag}(\pi_1,\ldots,\pi_V)-\pi\pi^\top.
\end{equation}

The use of only the first $V$ categories is important. If all $V+1$
categories were included, the covariance matrix would be singular because
the category counts sum to the total number of words. In the reduced
$V$-dimensional representation used here, $\Sigma_p$ is positive definite
whenever
\[
\pi_1,\ldots,\pi_V,\pi_{V+1}>0.
\]
Its inverse is
\[
\Sigma_p^{-1}
=
\operatorname{diag}\left(
\frac{1}{\pi_1},\ldots,\frac{1}{\pi_V}
\right)
+
\frac{1}{\pi_{V+1}}\mathbf{1}_V\mathbf{1}_V^\top.
\]

We allow the monthly word totals to be deterministic or random. In the
latter case, we assume that they are independent of the sequence
$\{Z_\ell\}_{\ell\geq1}$.

\begin{theorem}\label{thm:fclt}
Suppose that, as $T\to\infty$,
\[
M_T\xrightarrow{\PP}\infty
\]
and that there exists a continuous, nondecreasing function
$\Lambda:[0,1]\to[0,1]$, satisfying $\Lambda(0)=0$ and $\Lambda(1)=1$,
such that
\begin{equation}\label{eq:time-change}
\sup_{t\in[0,1]}
\left|
\frac{M_{\lfloor Tt\rfloor}}{M_T}-\Lambda(t)
\right|
\xrightarrow{\PP}0.
\end{equation}
Define
\[
D_T(t)
=
\frac{1}{\sqrt{M_T}}
\left(
C_{\lfloor Tt\rfloor}
-
\frac{M_{\lfloor Tt\rfloor}}{M_T}C_T
\right),
\qquad t\in[0,1].
\]
Then, in $D([0,1],\R^V)$,
\[
D_T
\Rightarrow
\Sigma_p^{1/2}B_0\bigl(\Lambda(\,\cdot\,)\bigr),
\]
where
\[
B_0(u)=W(u)-uW(1),
\qquad 0\leq u\leq1,
\]
and $W$ is a standard $V$-dimensional Brownian motion.
\end{theorem}

\begin{prf}
For $n\geq1$ and $u\in[0,1]$, define
\[
G_n(u)
=
\frac{1}{\sqrt{n}}
\sum_{\ell=1}^{\lfloor nu\rfloor}(Z_\ell-\pi).
\]
By the multivariate functional central limit theorem,
\[
G_n\Rightarrow\Sigma_p^{1/2}W
\]
in $D([0,1],\R^V)$. Since $M_T\to\infty$ in probability and, whenever
$M_T$ is random, it is independent of $\{Z_\ell\}_{\ell\geq1}$, the
randomly indexed version also satisfies
\[
G_{M_T}\Rightarrow\Sigma_p^{1/2}W.
\]

Set
\[
u_T(t)=\frac{M_{\lfloor Tt\rfloor}}{M_T}.
\]
By \eqref{eq:time-change},
\[
\lVert u_T-\Lambda\rVert_\infty\xrightarrow{\PP}0.
\]
Moreover, since $M_{\lfloor Tt\rfloor}$ is integer-valued,
\[
\left\lfloor M_Tu_T(t)\right\rfloor
=
M_{\lfloor Tt\rfloor}.
\]
Consequently,
\[
\frac{
C_{\lfloor Tt\rfloor}-M_{\lfloor Tt\rfloor}\pi
}{\sqrt{M_T}}
=
G_{M_T}\bigl(u_T(t)\bigr),
\]
and
\[
\frac{C_T-M_T\pi}{\sqrt{M_T}}=G_{M_T}(1).
\]
Therefore, the following identity is exact:
\begin{align*}
D_T(t)
&=
\frac{
C_{\lfloor Tt\rfloor}-M_{\lfloor Tt\rfloor}\pi
}{\sqrt{M_T}}
-
u_T(t)
\frac{C_T-M_T\pi}{\sqrt{M_T}}
\\
&=
G_{M_T}\bigl(u_T(t)\bigr)-u_T(t)G_{M_T}(1).
\end{align*}
The limiting Brownian motion has continuous sample paths, while $\Lambda$
is continuous by assumption. Hence, the continuous mapping theorem gives
\begin{align*}
D_T(t)
&\Rightarrow
\Sigma_p^{1/2}
\left[
W\bigl(\Lambda(t)\bigr)-\Lambda(t)W(1)
\right]
\\
&=
\Sigma_p^{1/2}B_0\bigl(\Lambda(t)\bigr),
\end{align*}
which proves the assertion.
\end{prf}

\begin{remark}\label{rem:no-dominating-month}
The condition that $\Lambda:[0,1]\to[0,1]$ is continuous is equivalent to the condition that no single month dominates,
\begin{equation}\label{eq:no-dominating-month}
\max_{1\leq i\leq T}\frac{m_i}{M_T}
\xrightarrow{\PP}0.
\end{equation}
This condition is natural because it ensures that the largest jump of the cumulative
time-change process is asymptotically negligible. However,
\eqref{eq:no-dominating-month} alone is not sufficient to identify the
limiting time change. Condition \eqref{eq:time-change}, or another set of
assumptions implying it, is also required.
\end{remark}

Since
\[
\widehat{\pi}=\frac{C_T}{M_T}\xrightarrow{\PP}\pi,
\]
we also have
\[
\widehat{\Sigma}_p\xrightarrow{\PP}\Sigma_p
\qquad\text{and}\qquad
\widehat{\Sigma}_p^{-1}\xrightarrow{\PP}\Sigma_p^{-1},
\]
provided that $\pi_1,\ldots,\pi_{V+1}$ are strictly positive.

\begin{corollary}\label{cor:B-limit}
Assume the conditions of Theorem~\ref{thm:fclt} and suppose additionally
that
\[
0<\Lambda(\varepsilon)
\leq \Lambda(1-\varepsilon)<1.
\]
Then the statistic in \eqref{statB} satisfies
\[
B
\Rightarrow
\sup_{t\in[\varepsilon,1-\varepsilon]}
\frac{
\left\lVert B_0\bigl(\Lambda(t)\bigr)\right\rVert^2
}{
\Lambda(t)\bigl(1-\Lambda(t)\bigr)
}.
\]
Since $\Lambda$ is continuous and nondecreasing, this limit can equivalently
be written as
\[
\sup_{u\in[\Lambda(\varepsilon),\Lambda(1-\varepsilon)]}
\frac{\lVert B_0(u)\rVert^2}{u(1-u)}.
\]
\end{corollary}

\begin{prf}
For $k=\lfloor Tt\rfloor$, write
\(
u_T(t)=\frac{M_k}{M_T}.
\)
By \eqref{eq:dk},
\[
\frac{d_k^\top\widehat{\Sigma}_p^{-1}d_k}
{M_k(M_T-M_k)/M_T}
=
\frac{
D_T(t)^\top\widehat{\Sigma}_p^{-1}D_T(t)
}{
u_T(t)(1-u_T(t))
}.
\]
The result now follows from Theorem~\ref{thm:fclt}, the consistency of
$\widehat{\Sigma}_p^{-1}$, and the continuous mapping theorem.
\end{prf}

\begin{corollary}[Cumulative-word trimming]
\label{cor:cumulative_word_trimming}
Assume the conditions of Theorem~\ref{thm:fclt}, with
$\pi_1,\ldots,\pi_V,\pi_{V+1}>0$.

For a fixed $\varepsilon\in(0,1/2)$, define
\[
B^\star
=
\max_{\substack{1\leq k\leq T-1\\
                \varepsilon\leq M_k/M_T\leq 1-\varepsilon}}
\frac{
    d_k^\top\widehat{\Sigma}_p^{-1}d_k
}{
    M_k(M_T-M_k)/M_T
}.
\]
Then, as $T\to\infty$,
\[
B^\star
\Rightarrow
\sup_{u\in[\varepsilon,1-\varepsilon]}
\frac{\|B_0(u)\|^2}{u(1-u)},
\]
where $B_0$ is a standard $V$-dimensional Brownian bridge.
\end{corollary}

\begin{proof}
Let $G_n$ be the partial-sum process introduced in the proof
of Theorem~\ref{thm:fclt}, and define
\[
H_T(u)=G_{M_T}(u)-uG_{M_T}(1),
\qquad u\in[0,1].
\]
The randomly indexed functional central limit theorem and
the continuous mapping theorem yield
\[
H_T\Rightarrow\Sigma_p^{1/2}B_0
\qquad\text{in }D([0,1],\mathbb{R}^V).
\]
Together with
$\widehat{\Sigma}_p^{-1}\xrightarrow{P}\Sigma_p^{-1}$,
this implies
\[
Q_T(u)
:=
\frac{
    H_T(u)^\top\widehat{\Sigma}_p^{-1}H_T(u)
}{
    u(1-u)
}
\Rightarrow
\frac{\|B_0(u)\|^2}{u(1-u)}
\]
on $[\varepsilon,1-\varepsilon]$, with a continuous
limiting process.

At each cumulative-word fraction $u=M_k/M_T$, we have
\[
H_T(M_k/M_T)=\frac{d_k}{\sqrt{M_T}}.
\]
Consequently, $B^\star$ is the maximum of $Q_T$ over the
grid
\[
\mathcal{U}_T
=
\left\{
\frac{M_k}{M_T}:
1\leq k\leq T-1,\quad
\varepsilon\leq\frac{M_k}{M_T}\leq1-\varepsilon
\right\}.
\]
Since
\[
\max_{1\leq i\leq T}\frac{m_i}{M_T}
\xrightarrow{P}0,
\]
this grid becomes dense in
$[\varepsilon,1-\varepsilon]$ in probability.
The continuity of the limiting process therefore allows
the maximum over $\mathcal{U}_T$ to be replaced
asymptotically by the supremum over the entire interval.
This proves the assertion.
\end{proof}

We further use $B^\star$ in the simulation study and the real-world example.

\begin{remark}
The statistics $B$ and $B^\star$ differ only in the set
of admissible change-point locations. The former uses
calendar-time trimming, whereas the latter uses trimming
on the cumulative-word scale.

Unlike the limiting distribution in Corollary~\ref{cor:B-limit}, the
limiting distribution of $B^\star$ does not depend on
the unknown time-change function $\Lambda$. For fixed
$V$ and $\varepsilon$, it is also independent of the
word-category probabilities. Thus, under the multinomial
null model, $B^\star$ admits an asymptotic calibration
that does not require estimation of $\Lambda$, even when
the monthly word totals vary substantially over time.

If $\Lambda(t)=t$, the calendar-time and cumulative-word
trimming schemes yield the same limiting null distribution.
\end{remark}

\begin{remark}\label{rem:dirichlet}
The preceding result is derived under the multinomial null model. For data
generated from a Dirichlet or Dirichlet--multinomial model, words within the
same document are generally dependent and the covariance matrix in
\eqref{eq:sigma-p} is no longer the correct covariance matrix of the
aggregated observations. Consequently, the null distribution will in
general depend on the Dirichlet concentration parameters unless the
statistic is standardized by an appropriate covariance estimator or
calibrated by a bootstrap procedure that preserves the document-level
dependence. Therefore, Theorem~\ref{thm:fclt} and Corollary~\ref{cor:cumulative_word_trimming}  directly justify statistics
$B$ and $B^\star$ only under the multinomial model; the robustness under Dirichlet
alternatives should be investigated by simulation or treated using a
separate document-level asymptotic argument.
\end{remark}

\section{Simulation study}\label{sec4}

The primary goal
is to explore the limitations of applying the asymptotic null distribution of  $B^\star$ given in Section \ref{sec3} and to offer practical recommendations and alternative approaches for approximating p-values.


To reach the goal, we consider the following settings:

\begin{enumerate}
    \item Keyword count vectors follow  multinomial $Mult(n_{ij},\pi)$, while the number of documents per month  is generated from a {shifted Poisson $2 + \mathcal{P}(32)$. The shift ensures at least two documents per month and substantially speeds up the simulation code by eliminating the need to handle months with zero or one document separately.}; 
     \item Keyword count vectors follow  multinomial $Mult(n_{ij},\pi)$, while the number of documents per month  is generated from a shifted Poisson $2 + \mathcal{P}(\lambda_i)$, where $\lambda^*_i=\frac{1}{4}32^{\frac{i-1}{99}},$ and $\lambda_i=\frac{32T}{\sum_{j=1}^T\lambda^*_j}\lambda^*_i$. This selection of parameters is  taken to model the inhomogeneity of the documents across time, having the average number of documents equal to the previous homogeneous scenario. With this setup, the average monthly number of documents approximately doubles every two years;
     \item {Keyword count vectors follow a Dirichlet mixture of multinomial  distributions. Each document is generated according to a multinomial $Mult(n_{ij},\pi_{ij})$ distribution, while the distribution $\pi_{ij}$ is a random variable, calculated as the convex combination of distributions $\pi^1,\dots,\pi^K$, where the mixture values are drawn from the symmetrical Dirichlet $Dir(\alpha)$. The number of documents per month is generated in the same way as in setting 1. }
     \item { Keyword count vectors follow the same distribution as in setting 3, while the number of documents per month is generated according to the same rules as in setting 2.}
\end{enumerate}
In all scenarios, we consider $T=100$ months and the number of keywords tracked $k\in\{3,5,10\}.$ The locations of the change-points considered are $\tau=10..90(10)$. In all scenarios, the length of each document is modeled as $\mathcal{P}(150)$. More details about the model parameters for different settings are given in Appendix \ref{appA}. In addition, the same settings have been studied with different average monthly documents and the results of those simulations are given in the Appendix \ref{appB}.

We consider two alternative resampling methods for calculating p-values. The Algorithm \ref{parametric} is a classical parametric bootstrap for multinomial models, while the nonparametric Algorithm \ref{nonparametric} adapts the classical permutation approach.

\begin{algorithm}
  \caption{Parametric bootstrap algorithm assuming multinomial distribution}\label{parametric}
  \begin{algorithmic}[1]
    \State Start with the sample $\boldsymbol{X}=\{X_{ij}\},\; i=1..T, j=1..N_i;$ and calculate change-point test statistic $A$;
    \State Estimate $\pi=(\pi_1,...,\pi_{V+1})^T$ with \eqref{ocenaPi};
    \State Construct a bootstrap sample  $\boldsymbol{X}^*=\{X^*_{ij}\},\; i=1..T,\; j=1..N_i;$ from $Mult(n_{ij},\hat{\pi})$.
    \State Based on the bootstrap sample calculate the change-point test statistic $A^*$;
    \State Repeat steps 3-4 $B$ times to obtain $A_{1}^{*}, A_{2}^{*}, \dots, A_{B}^{*}$;
    \State Reject the null hypothesis for the significance level $\alpha$ if $A$ is greater than the $(1 - \alpha)$-quantile of the empirical bootstrap distribution of $(A_{1}^{*}, A_{2}^{*}, \dots, A_{B}^{*})$.
  \end{algorithmic}
\end{algorithm}

\begin{algorithm}
  \caption{Nonparametric bootstrap algorithm}\label{nonparametric}
  \begin{algorithmic}[1]
    \State Start with the sample $\boldsymbol{X}=\{X_{ij}\},\; i=1..T, j=1..N_i;$ and calculate change-point test statistic $A$;
    \State Independently sample $N$ documents with replacement
    from the pooled collection. Each selected document
    retains its word-count vector and total word count;
    \State Partition the resampled sequence into consecutive
    monthly groups of sizes $N_1,\ldots,N_T$, keeping these
    group sizes fixed;
    \State Denote the resulting bootstrap sample by $\boldsymbol{X^\star}$;
    \State Based on the bootstrap sample calculate  the change-point test statistic $A^*$;
    \State Repeat steps 2-5 $B$ times to obtain $A_{1}^{*}, A_{2}^{*}, \dots, A_{B}^{*}$;
    \State Reject the null hypothesis for the significance level $\alpha$ if $A$ is greater than the $(1 - \alpha)$-quantile of the empirical bootstrap distribution of $(A_{1}^{*}, A_{2}^{*}, \dots, A_{B}^{*})$.
  \end{algorithmic}
\end{algorithm}

\begin{remark}
    Note that Algorithm \ref{nonparametric} implicitly assumes that the distribution of document lengths remains unchanged through the inspected period.
\end{remark}

\subsection{Competitor test}

The procedure proposed in \cite{preis2024change} also focuses
on a change-point in word proportions. However, since its application becomes infeasible for large number of documents, we focus on its adaptation for monthly aggregated word proportions. Its
implementation is substantially based on a normal approximation. Define
\[
Y_i
=
\frac{\sum_{j=1}^{N_i}X_{ij}}
{\sum_{j=1}^{N_i}n_{ij}},
\qquad
\overline{Y}
=
\frac{1}{T}\sum_{i=1}^T Y_i,
\]
and
\[
\widetilde{d}_k
=
\sum_{i=1}^k(Y_i-\overline{Y}).
\]
Let
\begin{align}\label{sigmatilde}
\widetilde{\Sigma}
=
\frac{1}{T-1}
\sum_{i=1}^T
(Y_i-\overline{Y})(Y_i-\overline{Y})^\top.
\end{align}

As a benchmark, we consider a monthly aggregated adaptation of the CUSUM procedure of \cite{preis2024change}. In this implementation, the covariance matrix is estimated from the complete sequence of monthly word-proportion vectors. We denote this statistic with $PS^{\star}$, i.e.

\[
PS^{\star}
=
\max_{\frac{k}{T}\in[\varepsilon,1-\varepsilon]}
\sqrt{
\frac{1}{T}
\widetilde{d}_k^\top
\widetilde{\Sigma}^{-1}
\widetilde{d}_k
}.
\]

\begin{algorithm}
  \caption{Parametric bootstrap algorithm assuming normal distribution}\label{parametricPS}
  \begin{algorithmic}[1]
    \State Start with the sample $\boldsymbol{X}=\{X_{ij}\},\; i=1..T, j=1..N_i;$ and derive $\boldsymbol{Y}=\{Y_i\},\; i=1..T$. 
    \State Calculate change-point test statistic $A$;
    \State Estimate $\mu=(\pi_1,...,\pi_{V})^T$ with $\bar{Y}$ and $\widetilde{\Sigma}$ with \eqref{sigmatilde}
    \State Construct a bootstrap sample  $\boldsymbol{Y}^*=\{Y^*_i\},\; i=1..T,\;$ from $\mathcal{N}(\hat{\mu},\widetilde{\Sigma})$.
    \State Based on the bootstrap sample calculate the change-point test statistic $A^*$;
    \State Repeat steps 3-4 $B$ times to obtain $A_{1}^{*}, A_{2}^{*}, \dots, A_{B}^{*}$;
    \State Reject the null hypothesis for the significance level $\alpha$ if $A$ is greater than the $(1 - \alpha)$-quantile of the empirical bootstrap distribution of $(A_{1}^{*}, A_{2}^{*}, \dots, A_{B}^{*})$.
  \end{algorithmic}
\end{algorithm}

\subsection{Results}

We run a Monte Carlo study to estimate the relevant performance of each test statistic in each setting. For the study, in each setting, we generate the monthly number of documents first, then we generate the length of each document. Afterwards, we perform 2000 Monte Carlo repetitions, each independently generating each documents' word frequencies. For practical purposes, we employ a warp-speed modifications of each presented bootstrap algorithm (see \cite{giacomini2013warp}).

The asymptotic distributions used in some scenarios are approximated in the following way: We sample 50000 trajectories of the limiting process, where each trajectory is approximated with 20000 discrete points in time. We then use the empirical quantiles of the limiting distribution, based on this sample. 

\subsubsection{Multinomial data generation}

We first examine the empirical Type I error rates of the proposed test statistic and the competitor in Settings 1 and 2, for different calibration methods, under the level of significance $0.05$.  
Table~\ref{tab:size}
presents the results for constant and exponentially
increasing document intensities.
We can see that the competitor, using the recommended normal bootstrap for calibration, significantly breaches the 0.05 threshold in the case of a growing trend in document prevalence, so we disregard calibration via the normal bootstrap in future analysis and focus on the competitor's test statistic performance using the same bootstrap procedures as in our proposed test.

\begin{table}
\centering
\renewcommand{\arraystretch}{1.08}
\begin{tabular}{llllr}
\toprule
Document intensity & Test statistic & Bootstrap / calibration & Reject rate \\
\midrule
\multicolumn{4}{l}{\textbf{Constant document intensity}} \\
Constant & Proposed  & Asymptotic    & 0.0385 \\
Constant & Proposed  & Multinomial   & 0.0530 \\
Constant & Proposed  & Nonparametric & 0.0495 \\
Constant & Competitor & Multinomial   & 0.0605 \\
Constant & Competitor & Nonparametric & 0.0490 \\
Constant & Competitor & Normal        & 0.0545 \\
\midrule
\multicolumn{4}{l}{\textbf{Growing document intensity}} \\
Growing & Proposed  & Asymptotic    & 0.0355 \\
Growing & Proposed  & Multinomial   & 0.0525 \\
Growing & Proposed  & Nonparametric & 0.0500 \\
Growing & Competitor & Multinomial   & 0.0675 \\
Growing & Competitor & Nonparametric & 0.0620 \\
Growing & Competitor & Normal        & 0.1350 \\
\bottomrule
\end{tabular}
\medskip
\caption{Type I error rates of the proposed and
reference tests under the multinomial null model for $V = 3$,
for constant and exponentially increasing document
intensity}\label{tab:size}
\end{table}

\begin{figure}[htbp]
    \centering

    \begin{subfigure}[t]{0.49\textwidth}
        \centering
        \includegraphics[width=\linewidth]
        {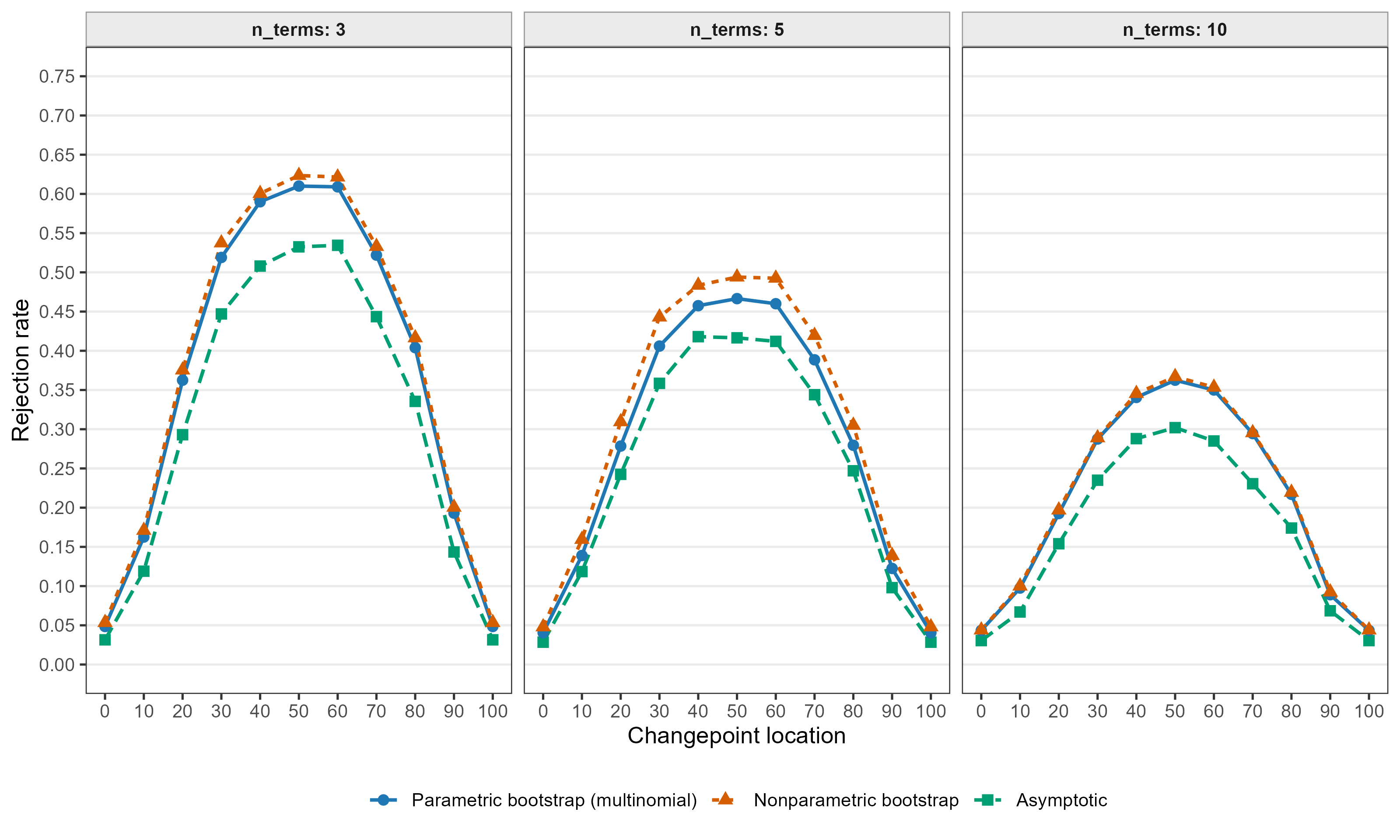}
        \caption{ Proposed test statistic, constant document intensity.}
    \end{subfigure}
    \hfill
    \begin{subfigure}[t]{0.49\textwidth}
        \centering
        \includegraphics[width=\linewidth]
        {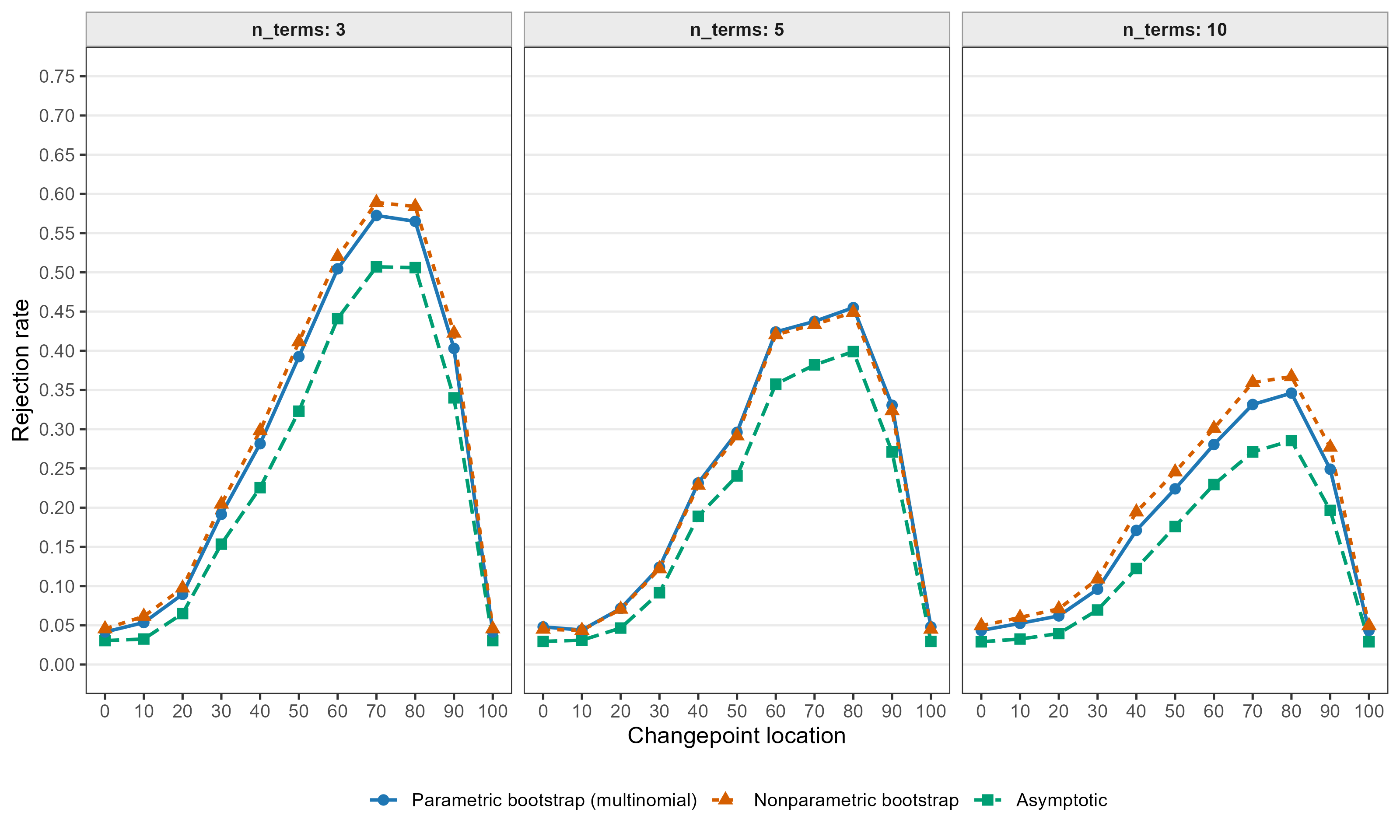}
        \caption{ Proposed test statistic, exponentially growing document intensity.}
    \end{subfigure}
\caption{Empirical rejection rates under multinomial data generation for the proposed test statistic under different calibration methods. Values at change-point locations 0 and 100 represent empirical test size.}
    \label{fig:multinomial_power_main}
\end{figure}

In Setting 1 (see Figure \ref{fig:multinomial_power_main}), we can see that the power for all three methods of calculating p-values reaches its maximum when the change-point is in the middle of the time interval. In setting 2 (see Figure \ref{fig:multinomial_power_main}), we see that the maximum power is reached when the change-point is around $80\%$ of the total time frame. At this point roughly $50\%$ of the total documents have been observed. The flexibility of the test statistic in this regard allows it to use the fact that more information is gathered in the later stages of the time frame.  

In both settings, we see that each approach to caluclating p-values is well calibrated, with the asymptotic approach being slightly conservative, thus achieving lower power. This is expected, as similar CUSUM-based finite-sample statistics have frequently been observed to yield smaller values than their asymptotic counterparts (see e.g.\cite{berkes2011asymptotics}). Calculating p-values from the asymptotic distribution is therefore only preferable when a very wide time frame is considered. Alternatively, other methods of assessing the distance to the asymptotic distribution could be implemented (\cite{yao1986asymptotic}).

\begin{figure}
    \begin{subfigure}[t]{0.49\textwidth}
        \centering
        \includegraphics[width=\linewidth]
        {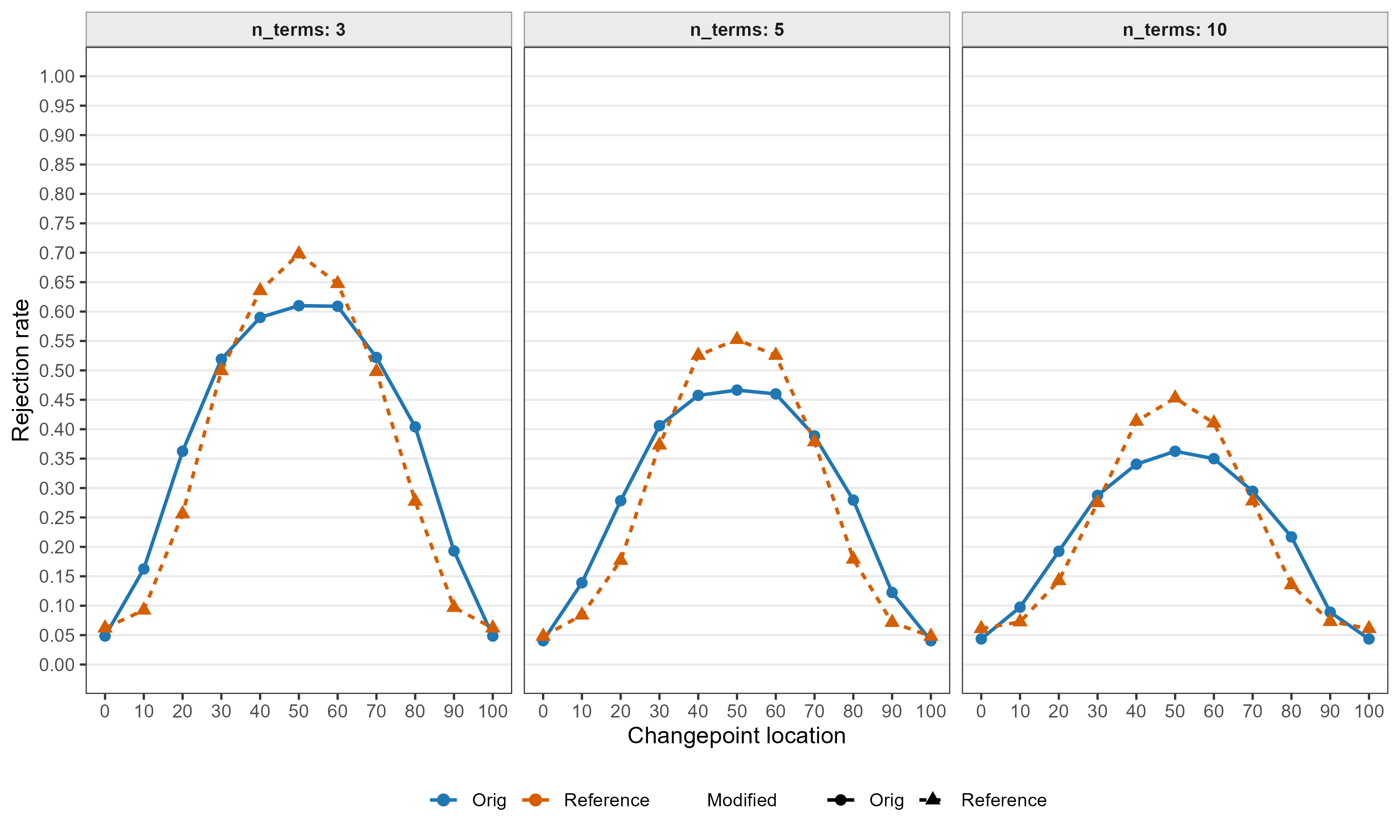}
        \caption{Parametric multinomial bootstrap, constant document intensity.}
    \end{subfigure}
    \hfill
    \begin{subfigure}[t]{0.49\textwidth}
        \centering
        \includegraphics[width=\linewidth]
        {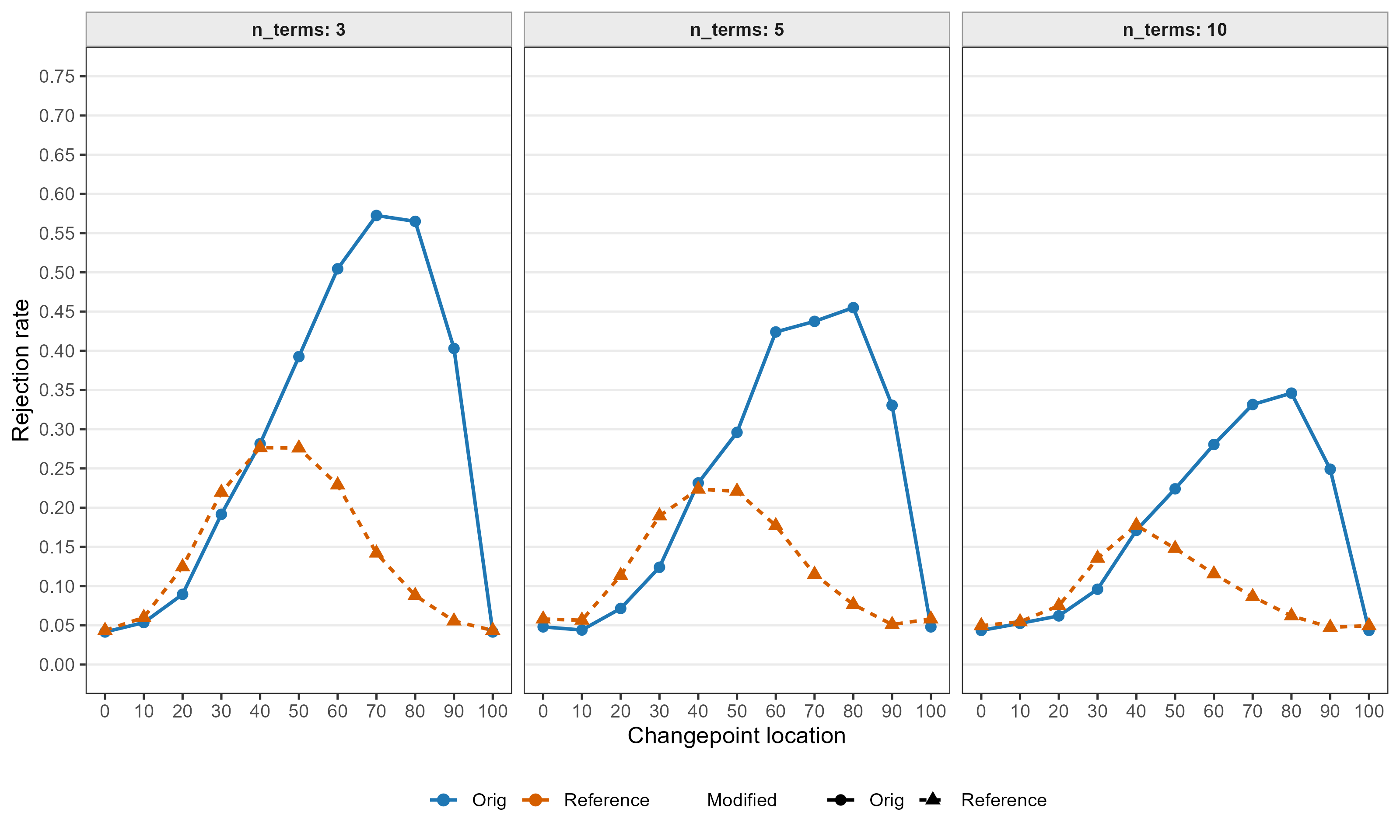}
        \caption{Parametric multinomial bootstrap, exponentially growing document intensity.}
    \end{subfigure}
    
    \vspace{0.5em}
    
    \begin{subfigure}[t]{0.49\textwidth}
        \centering
        \includegraphics[width=\linewidth]
        {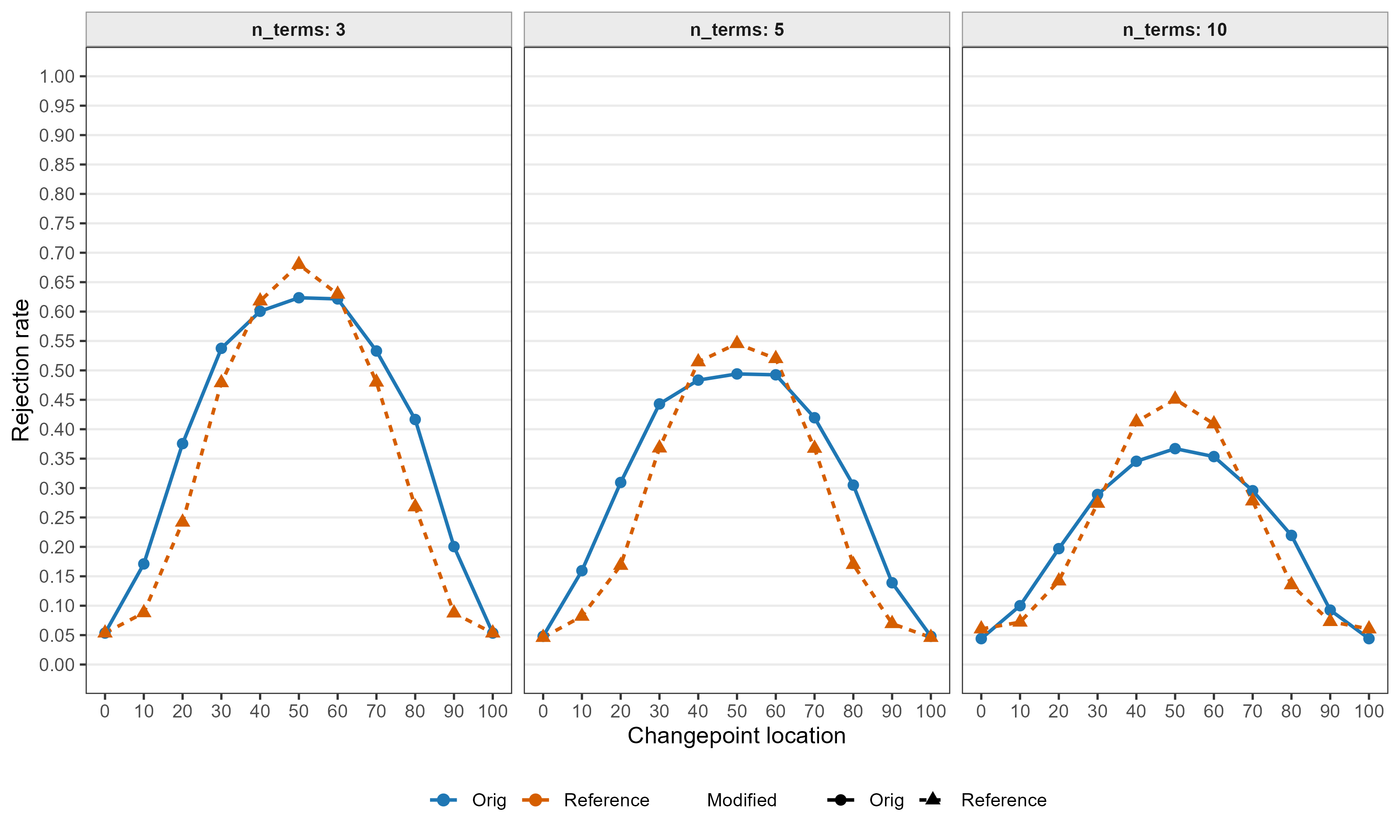}
        \caption{Nonparametric bootstrap, constant document intensity.}
    \end{subfigure}
    \hfill
    \begin{subfigure}[t]{0.49\textwidth}
        \centering
        \includegraphics[width=\linewidth]
        {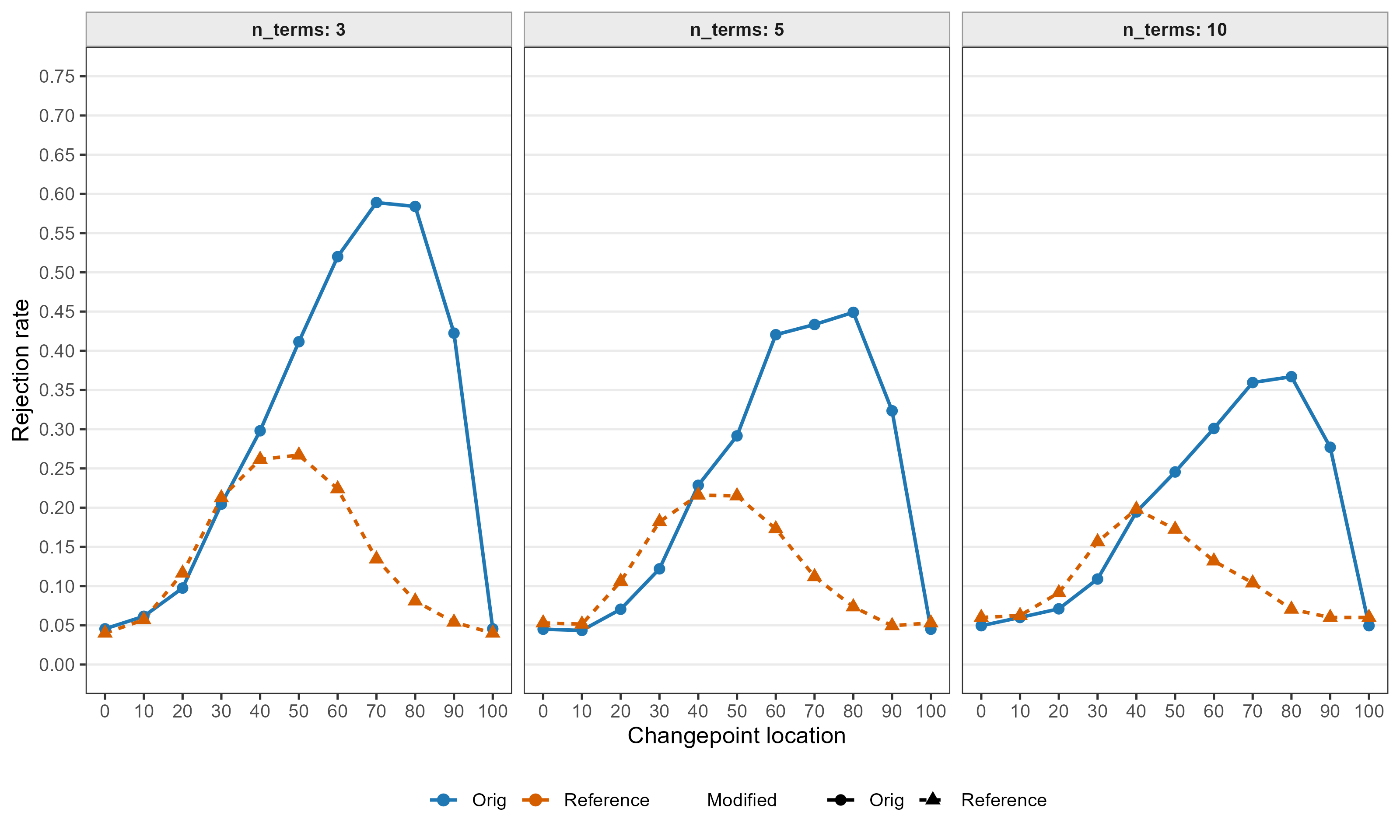}
        \caption{Nonparametric bootstrap, exponentially growing document intensity.}
    \end{subfigure}

    \caption{Empirical rejection rates under multinomial data generation - comparison to the competitor.
    Panels (a)--(b) compare the different test statistics under multinomial bootstrap calibration,
    while panels (c)--(d) compare them under the
    using the nonparametric bootstrap. Values at change-point locations 0 and 100 represent empirical test size.}
    \label{fig:multinomial_power_comp}
\end{figure}

Next we compare the power of the proposed test statistic and the competitor across different change-point locations, using the same bootstrap calibration (see Figure \ref{fig:multinomial_power_comp}). In Setting 1, whether we calibrate via the parametric or nonparametric bootstrap, we can see that the proposed test statistic and the competitor exhibit similar trends in power, but that the competitor performs slightly better when the change-point in the middle of the interval, but significantly worse as we approach the end points of the interval. This may be attributed in part to the variance normalization incorporated in the proposed statistic, which is absent from the competitor statistic. In Setting 2, the proposed statistic substantially outperforms the reference statistic across most change-point locations. Moreover, the proposed test maintains better control of the Type I error rate under increasing document intensity, as shown in Table \ref{tab:size}. We believe this is because the proposed test statistic properly leverages the extra information given in the later stages of the time frame.

\subsection{Dirichlet multinomial mixture data generation}

\begin{figure}[htbp]
    \centering

    \begin{subfigure}[t]{0.49\textwidth}
        \centering
        \includegraphics[width=\linewidth]
        {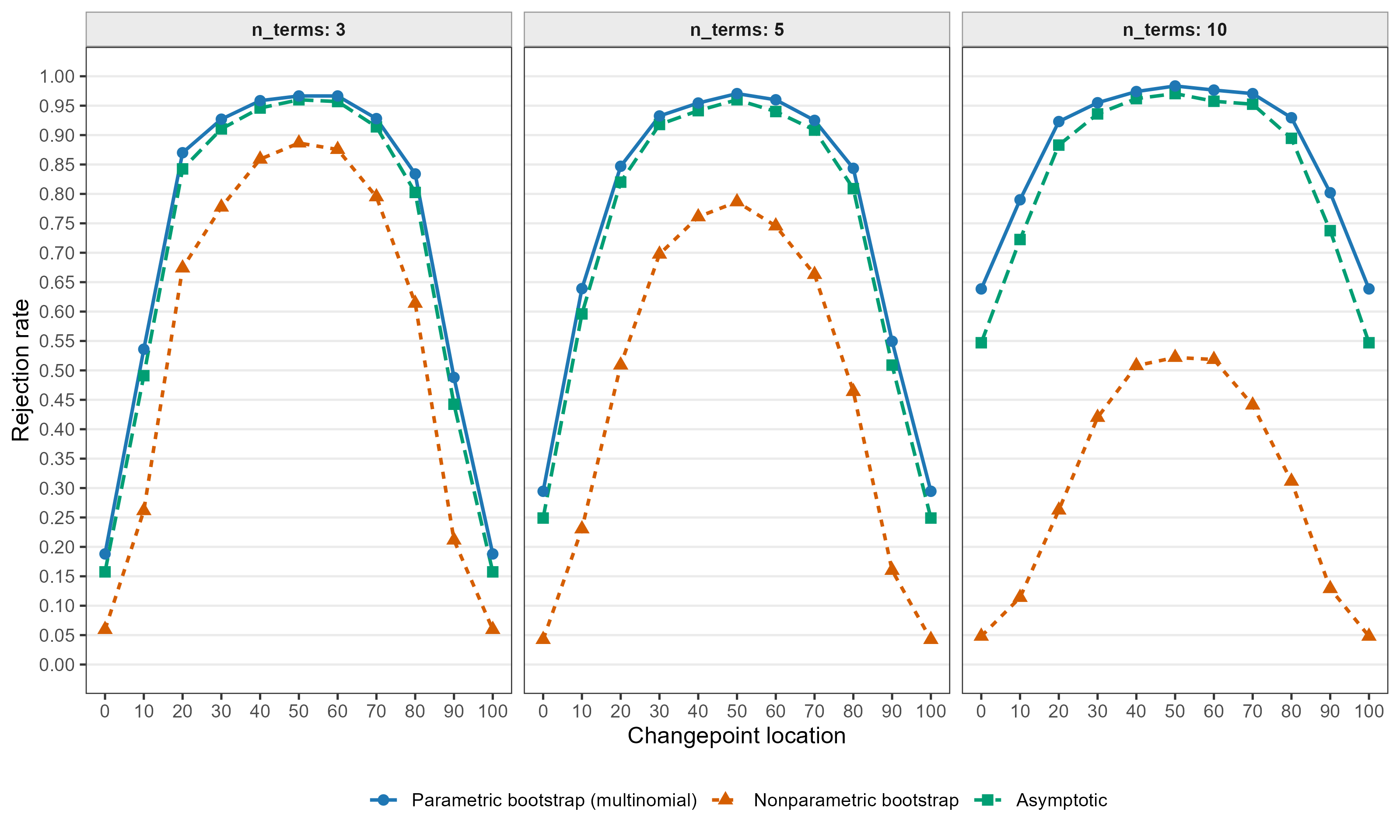}
        \caption{ Proposed test statistic, constant document intensity.}
    \end{subfigure}
    \hfill
    \begin{subfigure}[t]{0.49\textwidth}
        \centering
        \includegraphics[width=\linewidth]
        {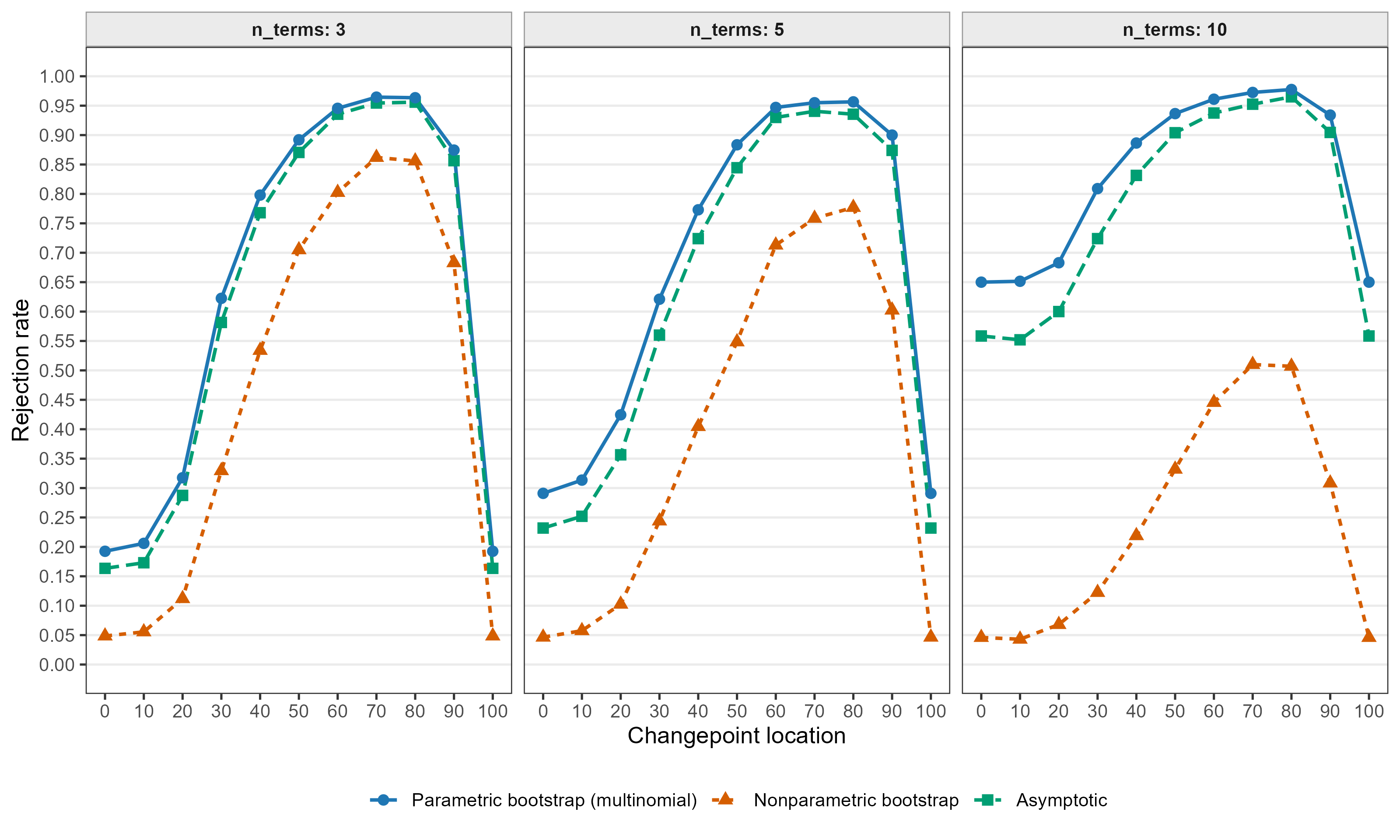}
        \caption{ Proposed test statistic, exponentially growing document intensity.}
    \end{subfigure}

    \vspace{0.5em}

    \begin{subfigure}[t]{0.49\textwidth}
        \centering
        \includegraphics[width=\linewidth]
{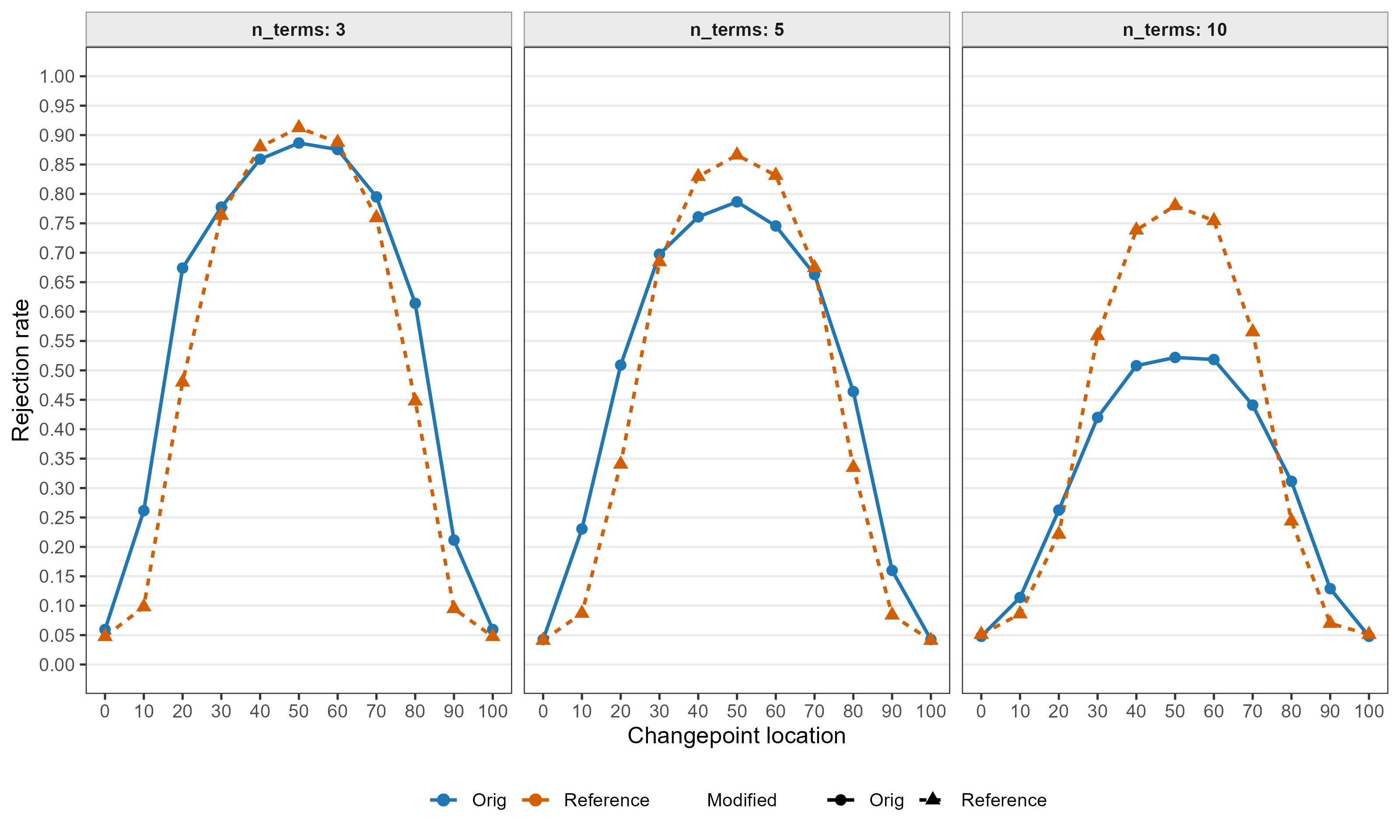}
        \caption{Constant document intensity.}
    \end{subfigure}
    \hfill
    \begin{subfigure}[t]{0.49\textwidth}
        \centering
        \includegraphics[width=\linewidth]
        {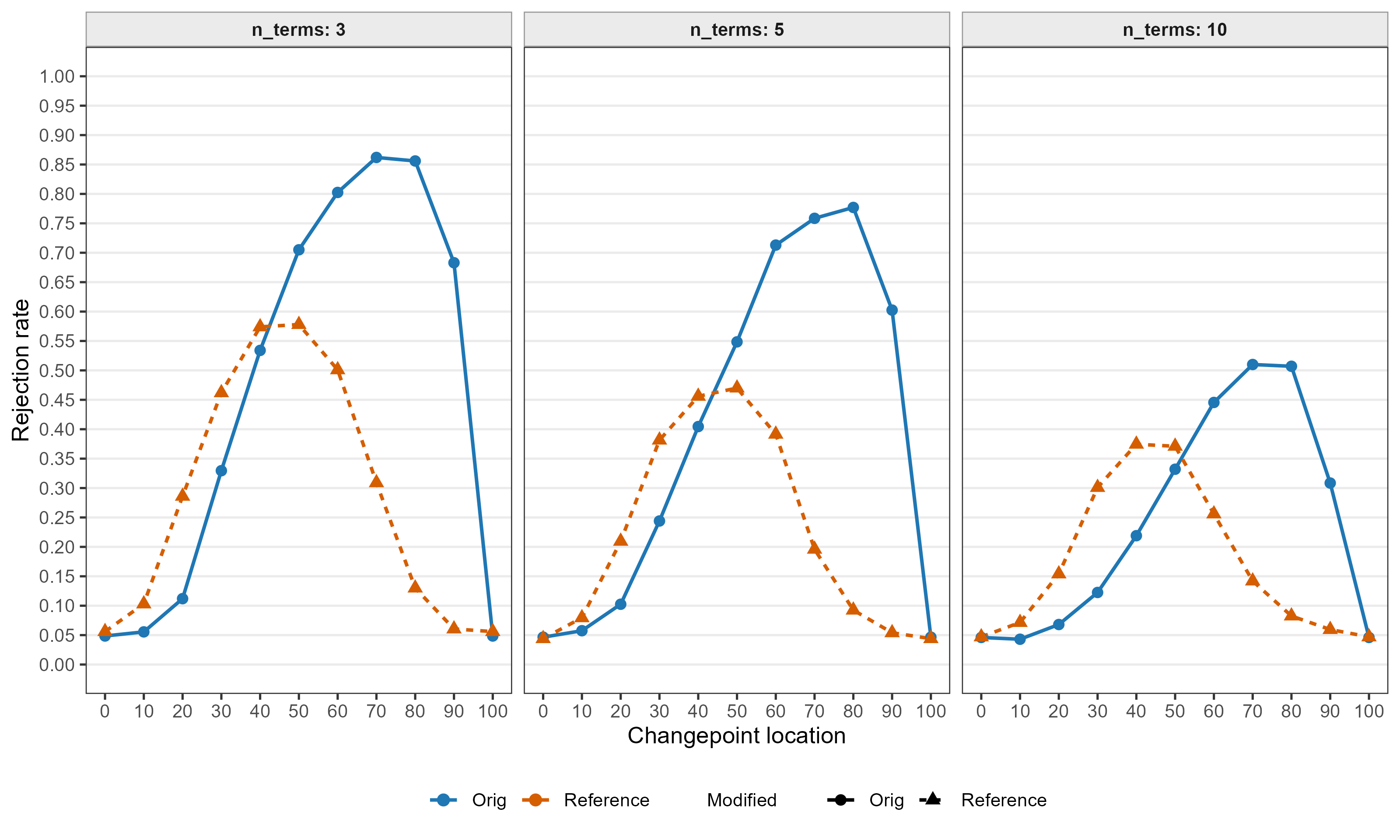}
        \caption{Exponentially growing document intensity.}
    \end{subfigure}
    
    \caption{Empirical rejection rates under Dirichlet mixture data generation.
    Panels (a)--(b) compare calibration methods for the original test statistic,
    while panels (c)--(d) compare the proposed test statistic with the competitor using the nonparametric bootstrap. Values at change-point locations 0 and 100 represent empirical test size.}
    \label{fig:dirichlet_power_main}
\end{figure}

Here we focus on Settings 3 and 4 (see Figure \ref{fig:dirichlet_power_main}). When comparing different methods of obtaining p-values, we immediately see that the type I error probability exceeds the set level of significance for all calibration methods bar the nonparametric bootstrap. We believe this happens due to the within-document dependence between word frequencies causing the test statistic to overestimate how much information it has gathered. 
By resampling entire documents, the nonparametric
bootstrap preserves the observed document-level
heterogeneity and provides more reliable calibration
in the considered settings. Therefore, in the
following comparison of the proposed and reference
statistics, we focus on the nonparametric bootstrap.

In this case, we see a similar trend where the proposed test statistic better finds proof at the ends of the time interval, and also adapts better to a growing trend in document frequency. We note that the difference is not as extreme as in the multinomial case. We also note that the competitor's power reduces at a lower rate when more terms are followed. We believe this to be the case because of the inconsistency of the multinomial parametric estimate of the covariance matrix. 

Based on these results, we suggest using the proposed statistic with fewer keywords, and with nonparametric p-value calibration, when we have reason to believe any divergence from the multinomial distribution is present.

\begin{figure}
    \centering
    \includegraphics[scale=0.6]{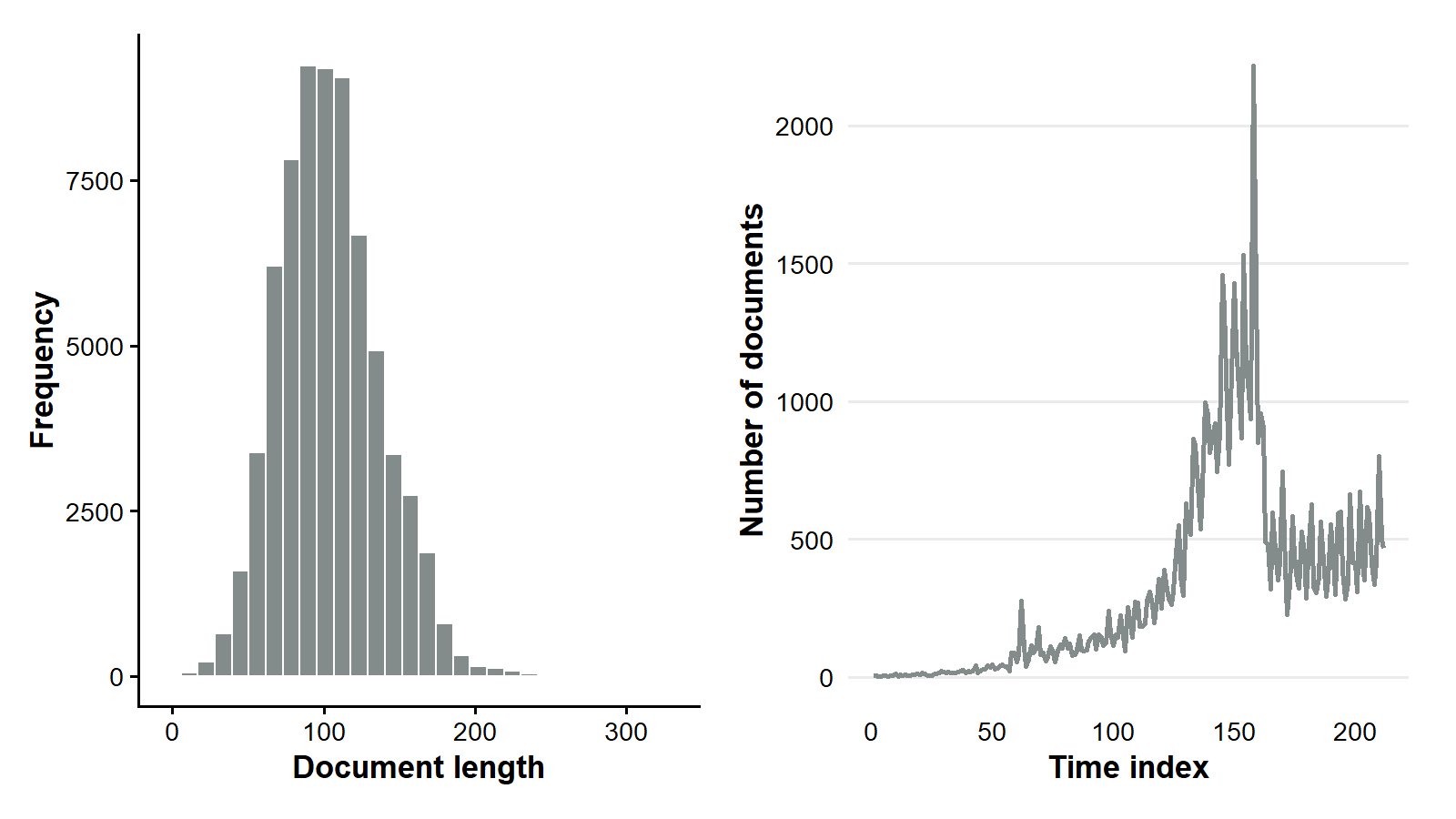}
    \caption{Histogram of abstract's length (left) and the evolution of the number of documents over time (right)}
    \label{fig:summaryarxiv}
\end{figure}

\begin{figure}
    \centering
    \includegraphics[scale=0.75]{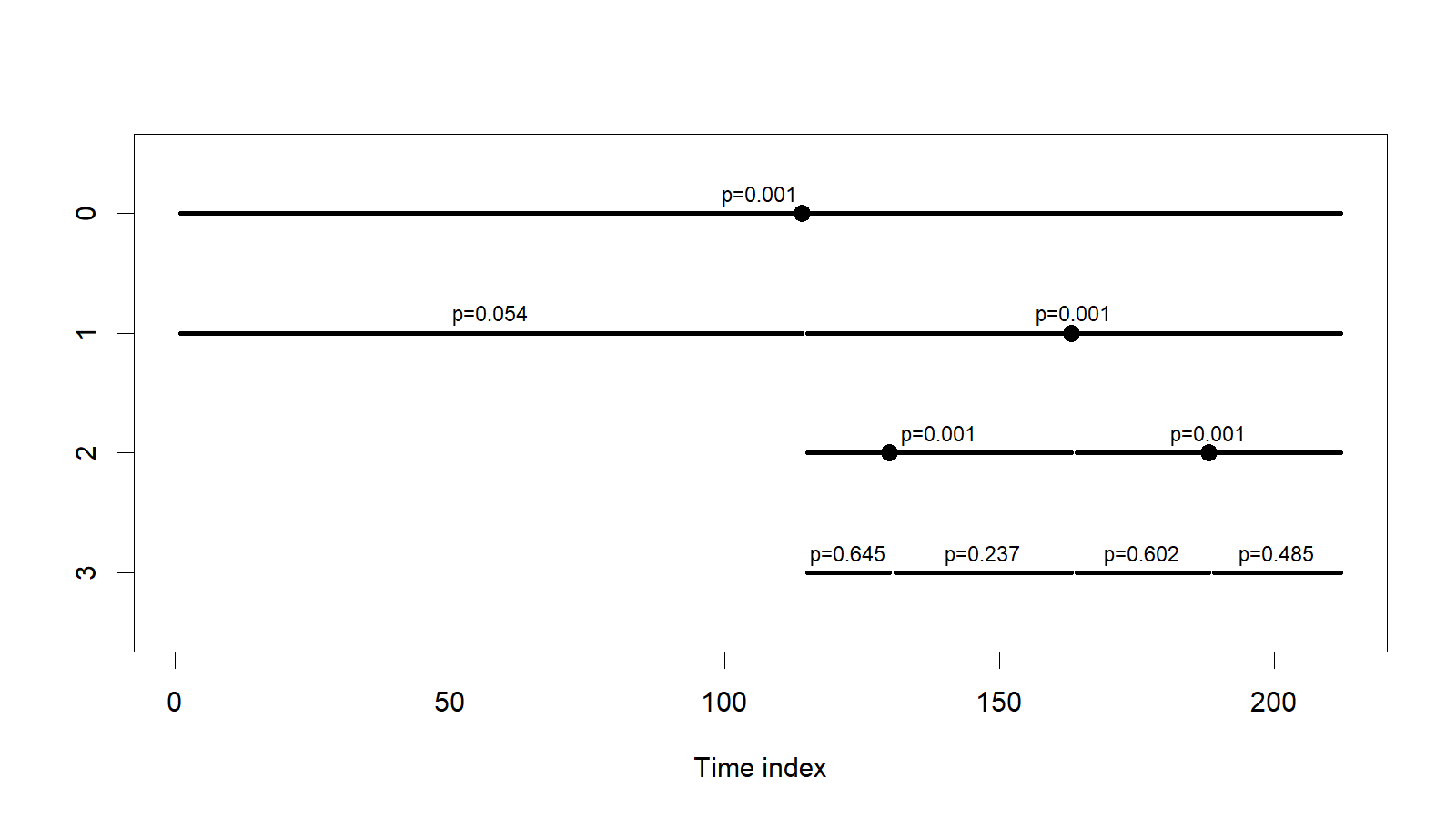}
    \caption{Binary segmentation multiple change-point detection}
    \label{fig:CParxiv}
\end{figure}


\section{Real data example: Arxiv abstract data}\label{sec5}
Here, we perform change-point analysis on the collection of abstracts from the arXiv category \emph{stat.ML}, covering the period from May 1, 2007, to January 1, 2025. The dataset contains 68,542 abstracts. We preprocess each text by removing punctuation, stop words and numbers and we lemmatize the remaining words. A summary of the data is presented in Figure \ref{fig:summaryarxiv}. The figure clearly indicates substantial variation in the number of documents over time. 

Moreover, there is insufficient justification for assuming a multinomial model for tracked keywords \emph{topic, cluster, learn} and, consequently, for relying on the corresponding asymptotic results. We therefore detect change-points using a nonparametric bootstrap procedure with $B=1000$ replicates. Applying binary segmentation, we recursively identify several change-points, with a significance level of 0.05 for each performed test, shown in Figure \ref{fig:CParxiv}.

The detected change-points can be interpreted in the context of major developments in statistical and machine learning research, although these associations should not be regarded as evidence of direct causality. The observation period begins in May 2007, when Statistics became a top-level archive on arXiv and Machine Learning (stat.ML) was introduced as one of its main categories \cite{ims2007arxiv}. The first change-point, in October 2016, falls within the period of rapid expansion of deep and representation learning, following major advances that substantially reshaped the machine-learning research landscape \cite{lecun2015deep}.
The second change-point, around February 2018, coincides with a period of increasing integration between statistics, machine learning, and the emerging data-science paradigm. Contemporary discussions explicitly described statistics as undergoing a transformation in response to big data, with supervised and unsupervised learning becoming increasingly prominent areas of statistical research \cite{donoho2017data,ley2018data}.
The third change-point, in November 2020, coincides with the growing prominence of large pretrained models, self-supervised approaches, and few-shot learning, exemplified by models such as GPT-3 \cite{brown2020language}. Finally, the change detected in December 2022 occurs immediately after the public release of ChatGPT on November 30, 2022 \cite{openai2022chatgpt}, which marked the beginning of an exceptionally rapid increase in research activity related to large language models, generative AI, and related learning paradigms. Taken together, these change-points appear to reflect broader shifts in the thematic structure and terminology of the statistical machine learning literature rather than isolated methodological developments.

{The p-values used for the decision-making were calculated based on the nonparametric bootstrap approach. This approach is the most robust, as seen in the previous simulation studies. Looking at the table below, we can see that the p-values based on the asymptotic distribution and the multinomial bootstrap are consistently much lower than the p-value obtained with the nonparametric bootstrap. This suggests that the multinomial model is not adequate for representing word frequencies in scientific abstracts, and instead the Dirichlet mixture model might be more appropriate.}
\begin{center}
\small
\begin{tabular}{rrrrrrr}
\toprule
Start & End & Depth & Nonparametric $p$ & CP location & Asymptotic $p$ & Parametric $p$ \\
\midrule
1   & 212 & 0 & 0.001 & 114 & 0.000 & 0.001 \\
1   & 114 & 1 & 0.054 & 55  & 0.000 & 0.001 \\
115 & 212 & 1 & 0.001 & 163 & 0.000 & 0.001 \\
115 & 163 & 2 & 0.001 & 130 & 0.000 & 0.001 \\
115 & 130 & 3 & 0.644 & 120 & 0.010 & 0.006 \\
131 & 163 & 3 & 0.237 & 152 & 0.000 & 0.001 \\
164 & 212 & 2 & 0.001 & 188 & 0.000 & 0.001 \\
164 & 188 & 3 & 0.601 & 170 & 0.011 & 0.008 \\
189 & 212 & 3 & 0.485 & 206 & 0.006 & 0.007 \\
\bottomrule
\end{tabular}
\end{center}

\section{Concluding  remarks and outlook}
 We have proposed a change-point detection
procedure for textual data based on word-frequency vectors.
Under the multinomial model, we have derived the asymptotic
null distribution of the proposed statistic, allowing for
nonuniform numbers of documents across time periods and
varying document lengths. In particular, the statistic
based on cumulative-word trimming has a limiting null
distribution that does not depend on the temporal
distribution of the observed textual information.

To improve finite-sample calibration, we have also
considered parametric and nonparametric bootstrap
procedures. While the parametric bootstrap relies on
the multinomial model, the nonparametric bootstrap
provides a more robust alternative in the presence
of document-level heterogeneity. Our simulation study
demonstrates the advantages of the proposed approach,
particularly when the number of documents varies
substantially over time or when the multinomial
assumption is violated. The application to arXiv
abstracts further illustrates its practical usefulness.

One of potential directions for future research is
to extend the proposed change-point methodology to
AI-based representations of textual data, such as
document embeddings obtained from transformer-based
language models (e.g., BERT \cite{devlin2019bert}  or Sentence-BERT \cite{reimers2019sentence}) and
representations derived from neural topic models (see e.g. \cite{dieng2020topic}).
Such extensions could enable the detection of changes
in the semantic structure of document collections,
beyond shifts in individual word frequencies. Another research direction entails modifying the procedure to better model the dependence between different words, which could potentially increase its power when there is a discrepancy between the multinomial model and the actual data generation mechanism.

\section*{Declaration}
AI tools were used for grammar and language editing,  and code formatting. No AI tools were used to generate data. The authors take full responsibility for the accuracy of the work presented.
\section*{Acknowledgments}
The work  is supported by the Ministry of Science, Technological Development and Innovations of the Republic of Serbia (the contract 451-03-33/2026-03/200104). The results are also obtained upon work from COST Action \emph{HiTEc-Text, functional and other high-dimensional data in econometrics: New models, methods, applications, CA21163}, supported by COST (European Cooperation in Science and Technology).

\bibliographystyle{abbrv}

\begin{appendices}
\section{Simulation settings}\label{appA}

{In settings 1 and 2, we use the following probabilities for the $k$ keywords under the null hypothesis. The values were selected to be of the same order of magnitude as the relative frequencies of keywords in scientific abstracts. Every probability vector of lower $k$ is also a subvector of higher $k$. This insures that the results across different values of $k$ are comparable.
\begin{flalign*}
k = 3: \quad &\pi = (0.0003, 0.0030, 0.0002) &&\\k = 5: \quad &\pi = (0.0003, 0.0030, 0.0002, 0.0008, 0.0012) &&\\k = 10: \quad &\pi = (0.0003, 0.0030, 0.0002, 0.0008, 0.0012, 0.0006, 0.0004, 0.0009, 0.0007, 0.0005) &&
\end{flalign*}
For these to properly represent distributions, the multinomial probability parameter is then constructed by adding another value to each probability vector, so that the sum of all probabilities is equal to 1. This last value should be considered as the joint probability of words other than the tracked keywords.

In settings 3 and 4, we generated documents from a Dirichlet mixture of $K = 10$ probability distributions, which are commonly interpreted as themes. In order to realistically model word frequencies, the mixture parameter $\alpha$ was set to 0.35. This was the value we obtained as the average estimate by fitting many LDA models on the data used in the real-world example. The probability distributions of the keywords in the 10 given themes under the null hypothesis are given below, for each value of $k$. The values were selected so that the themes are distinct between each other.
\[
\begin{aligned}
k=3:\quad \Pi
&=
\left(
\begin{array}{rrr}
0.045 & 0.005 & 0.060 \\
0.004 & 0.045 & 0.006 \\
0.006 & 0.005 & 0.045 \\
0.030 & 0.025 & 0.005 \\
0.005 & 0.030 & 0.025 \\
0.025 & 0.005 & 0.030 \\
0.018 & 0.018 & 0.018 \\
0.040 & 0.004 & 0.020 \\
0.020 & 0.040 & 0.004 \\
0.004 & 0.020 & 0.040
\end{array}
\right)
\\[1em]
k=5:\quad \Pi
&=
\left(
\begin{array}{rrrrr}
0.045 & 0.005 & 0.060 & 0.004 & 0.003 \\
0.004 & 0.045 & 0.006 & 0.005 & 0.004 \\
0.006 & 0.005 & 0.045 & 0.006 & 0.004 \\
0.030 & 0.025 & 0.005 & 0.050 & 0.006 \\
0.005 & 0.030 & 0.025 & 0.006 & 0.050 \\
0.025 & 0.005 & 0.030 & 0.006 & 0.004 \\
0.018 & 0.018 & 0.018 & 0.006 & 0.004 \\
0.040 & 0.004 & 0.020 & 0.025 & 0.004 \\
0.020 & 0.040 & 0.004 & 0.035 & 0.025 \\
0.004 & 0.020 & 0.040 & 0.018 & 0.018
\end{array}
\right)
\\[1em]
k=10:\quad \Pi
&=
\left(
\begin{array}{rrrrrrrrrr}
0.045 & 0.005 & 0.060 & 0.004 & 0.003 & 0.003 & 0.003 & 0.003 & 0.003 & 0.003 \\
0.004 & 0.045 & 0.006 & 0.005 & 0.004 & 0.003 & 0.003 & 0.003 & 0.003 & 0.003 \\
0.006 & 0.005 & 0.045 & 0.006 & 0.004 & 0.003 & 0.003 & 0.003 & 0.003 & 0.003 \\
0.030 & 0.025 & 0.005 & 0.050 & 0.006 & 0.006 & 0.003 & 0.003 & 0.003 & 0.003 \\
0.005 & 0.030 & 0.025 & 0.006 & 0.050 & 0.004 & 0.006 & 0.003 & 0.003 & 0.003 \\
0.025 & 0.005 & 0.030 & 0.006 & 0.004 & 0.055 & 0.004 & 0.006 & 0.003 & 0.003 \\
0.018 & 0.018 & 0.018 & 0.006 & 0.004 & 0.003 & 0.055 & 0.004 & 0.006 & 0.003 \\
0.040 & 0.004 & 0.020 & 0.025 & 0.004 & 0.006 & 0.003 & 0.055 & 0.004 & 0.006 \\
0.020 & 0.040 & 0.004 & 0.035 & 0.025 & 0.003 & 0.006 & 0.003 & 0.050 & 0.004 \\
0.004 & 0.020 & 0.040 & 0.018 & 0.018 & 0.003 & 0.003 & 0.006 & 0.003 & 0.050
\end{array}
\right)
\end{aligned}
\]
}

For the alternative hypothesis, in either case of data generation, we considered 1 moment in time after which the probability distribution changes. In the case of multinomial data generation, we switched the probability of the first and third keyword. In the case of Dirichlet mixture, we switched the probability of the first and third keyword of the first theme, keeping other themes the same and keeping the Dirichlet mixture parameter the same. 

\section{Additional simulation results}\label{appB}

\begin{figure}[htbp]
    \centering

    \begin{subfigure}[t]{0.49\textwidth}
        \centering
        \includegraphics[width=\linewidth]{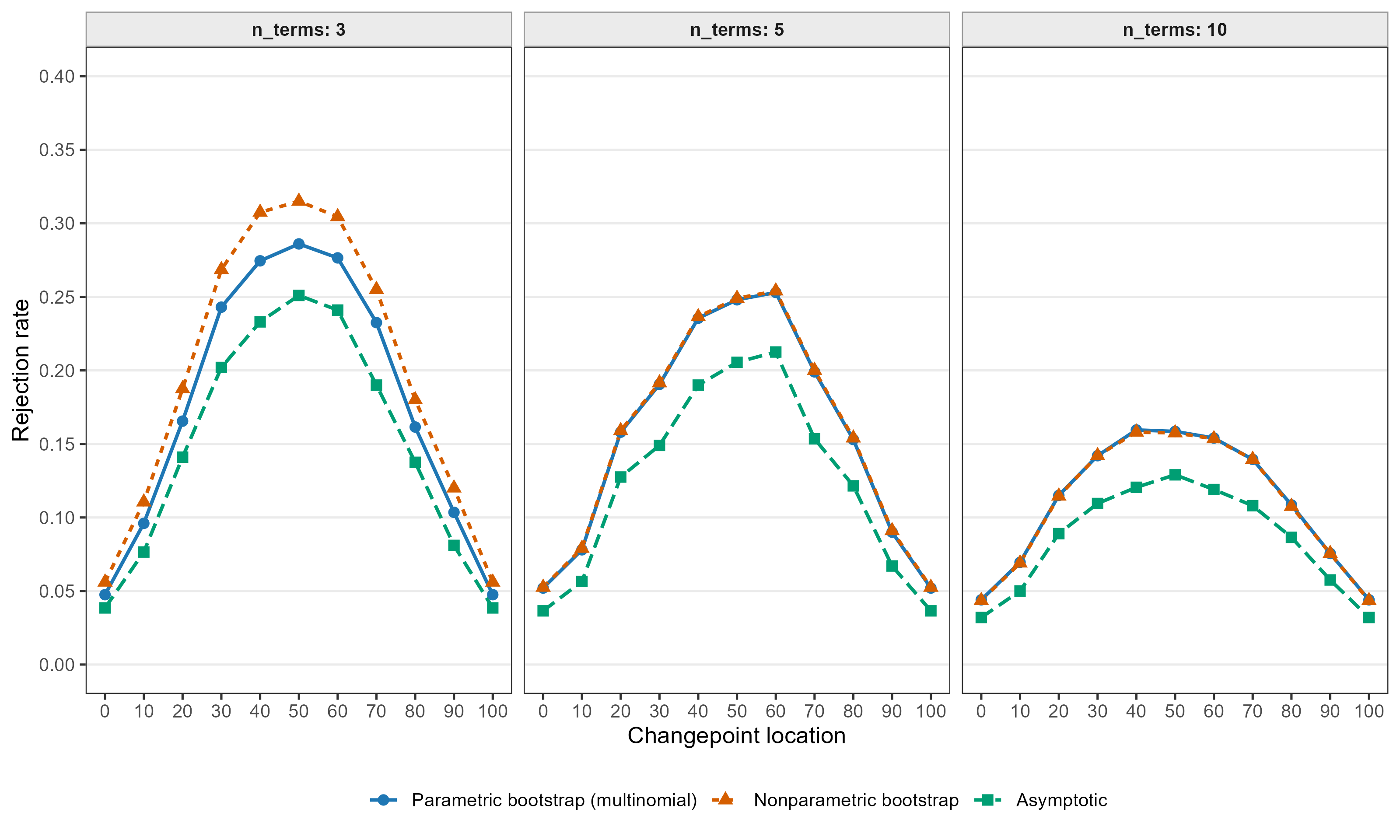}
        \caption{ Proposed test statistic, constant document intensity.}
    \end{subfigure}
    \hfill
    \begin{subfigure}[t]{0.49\textwidth}
        \centering
        \includegraphics[width=\linewidth]{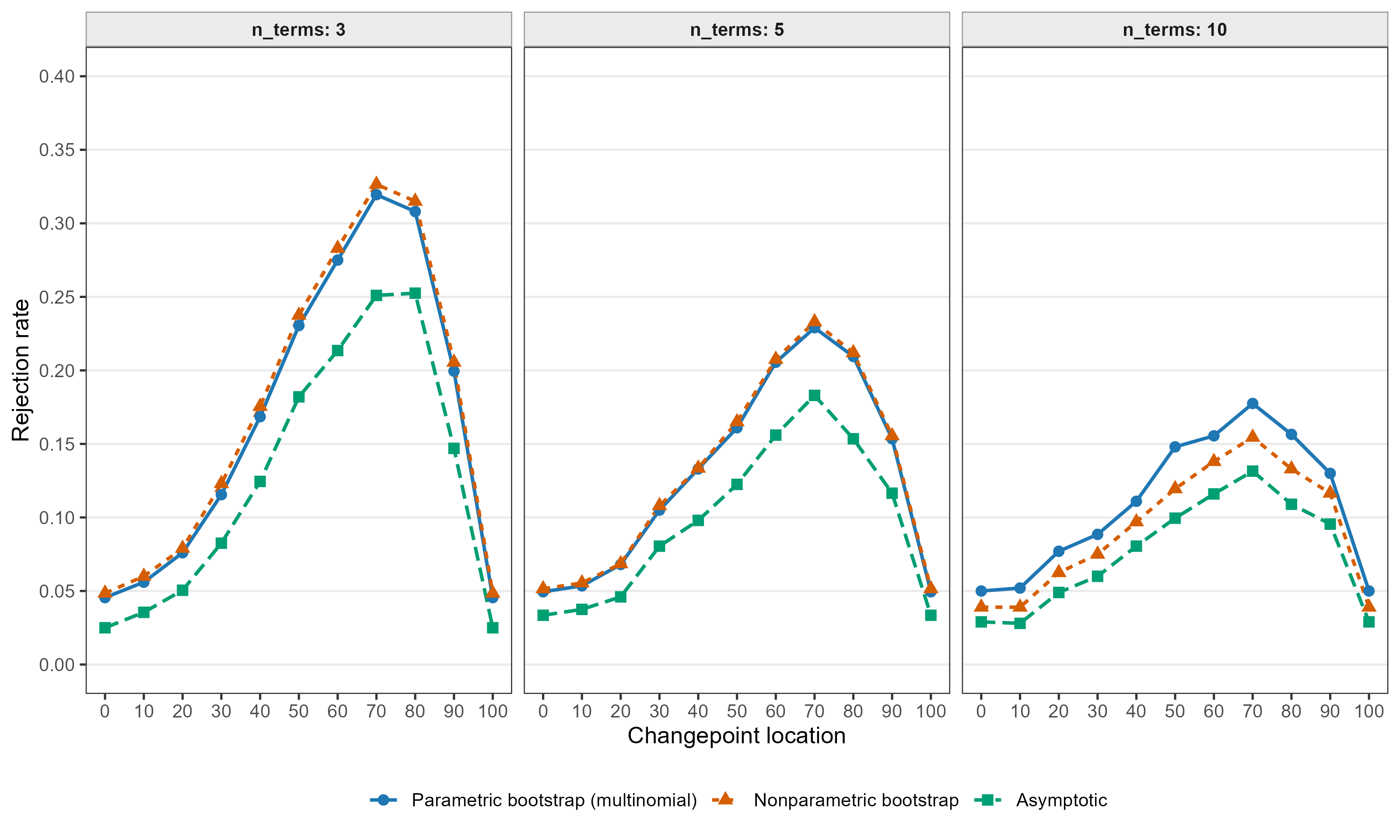}
        \caption{ Proposed test statistic, exponentially growing document intensity.}
    \end{subfigure}
\caption{Empirical rejection rates under multinomial data generation for the proposed test statistic under different calibration methods, with average monthly document intensity being equal 16.}
    \label{fig:multinomial_power_main_lambda16}
\end{figure}


\begin{figure}
    \begin{subfigure}[t]{0.49\textwidth}
        \centering
        \includegraphics[width=\linewidth]{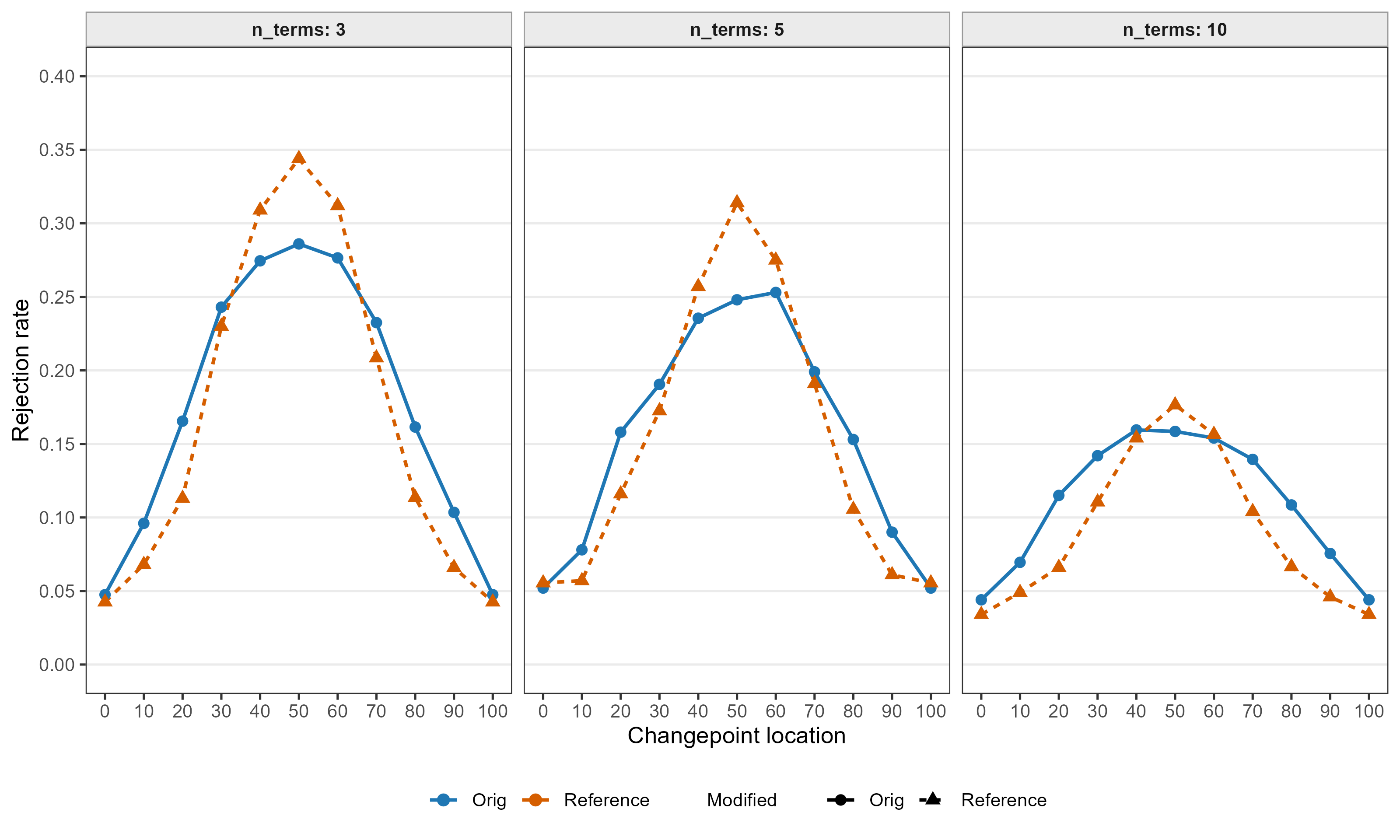}
        \caption{Parametric multinomial bootstrap, constant document intensity.}
    \end{subfigure}
    \hfill
    \begin{subfigure}[t]{0.49\textwidth}
        \centering
        \includegraphics[width=\linewidth]{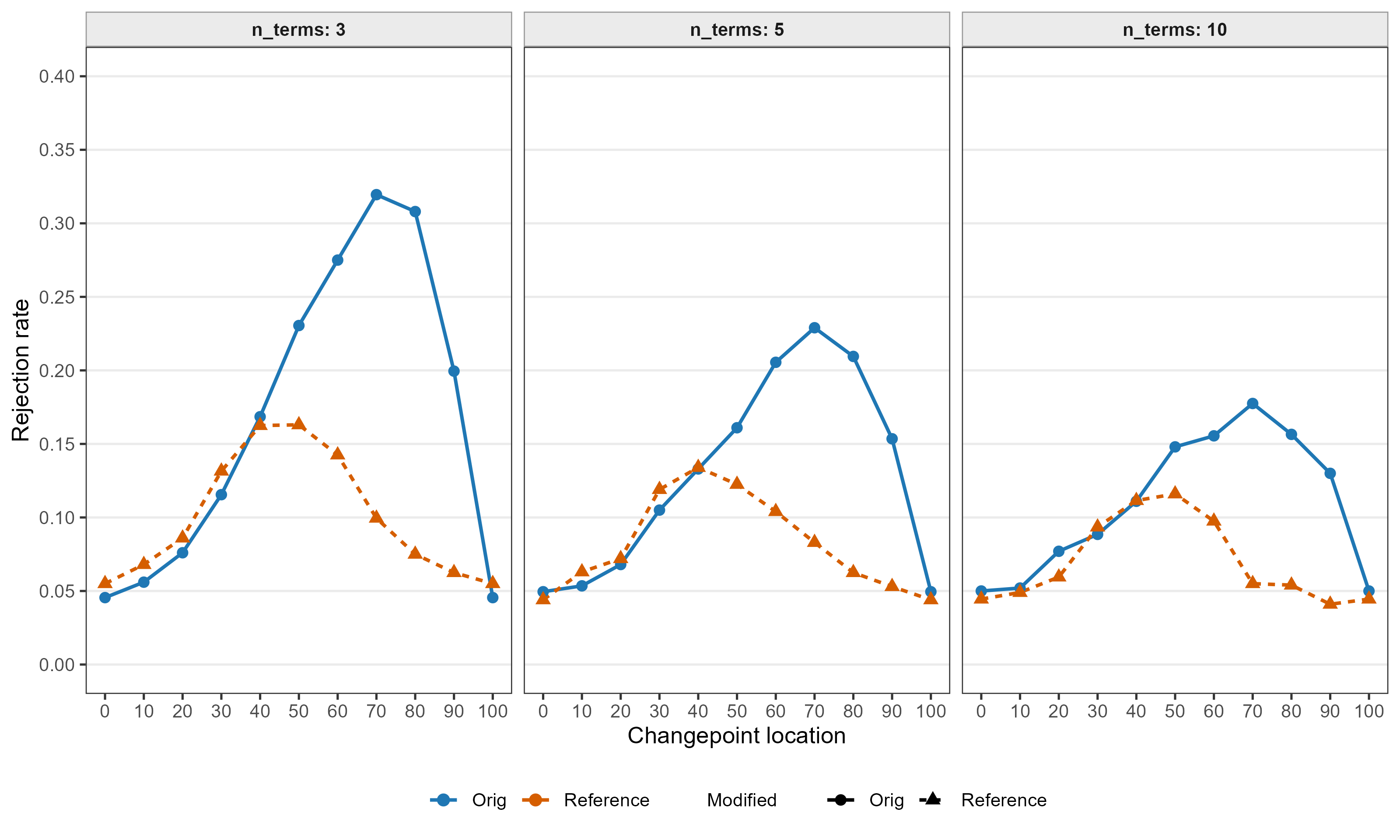}
        \caption{Parametric multinomial bootstrap, exponentially growing document intensity.}
    \end{subfigure}
    
    \vspace{0.5em}
    
    \begin{subfigure}[t]{0.49\textwidth}
        \centering
        \includegraphics[width=\linewidth]{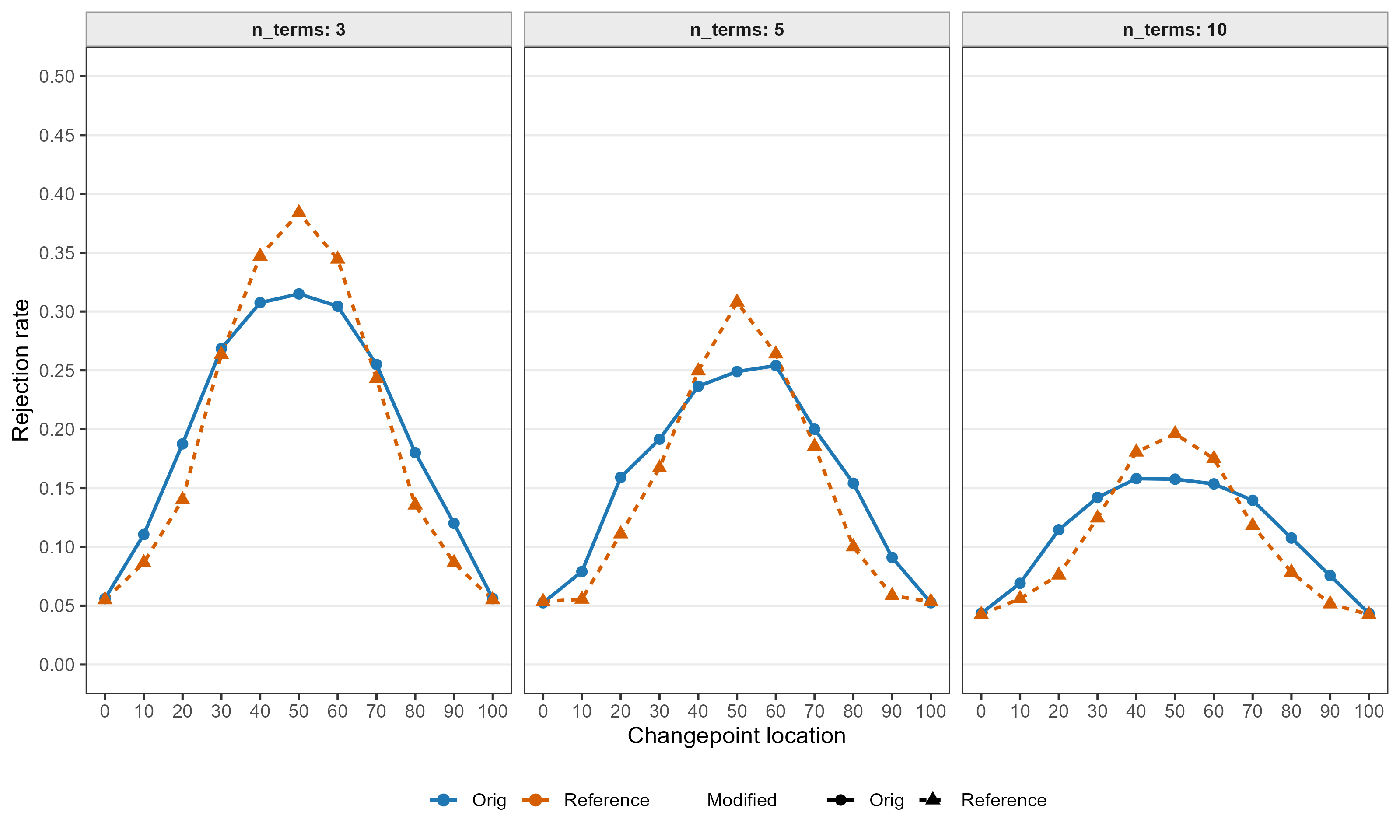}
        \caption{Nonparametric bootstrap, constant document intensity.}
    \end{subfigure}
    \hfill
    \begin{subfigure}[t]{0.49\textwidth}
        \centering
        \includegraphics[width=\linewidth]{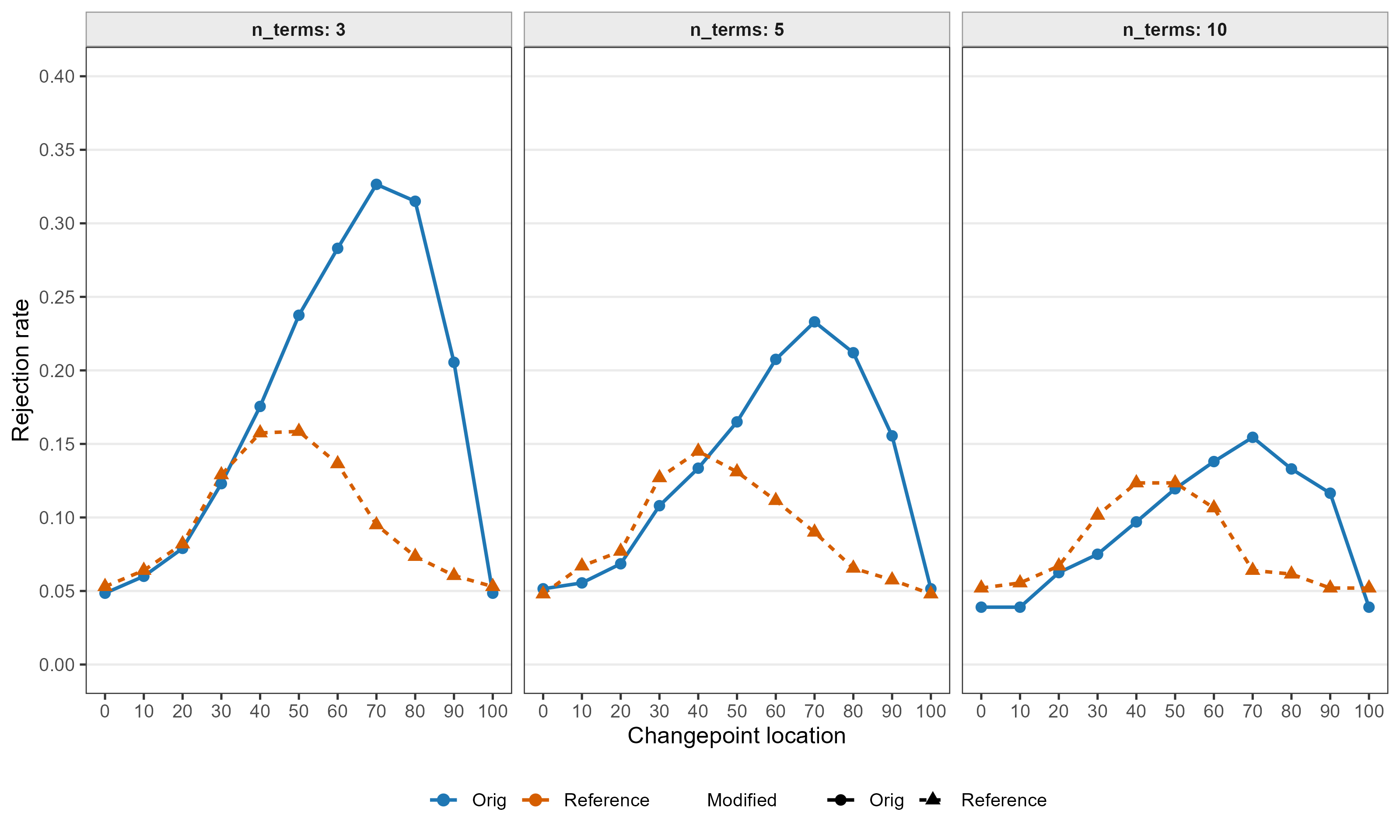}
        \caption{Nonparametric bootstrap, exponentially growing document intensity.}
    \end{subfigure}

    \caption{Empirical rejection rates under multinomial data generation - comparison to the competitor, with average monthly document intensity being equal 16.
    Panels (a)--(b) compare the different test statistics under multinomial bootstrap calibration,
    while panels (c)--(d) compare them under the
    using the nonparametric bootstrap.}
    \label{fig:multinomial_power_comp_lambda16}
\end{figure}

\begin{figure}[htbp]
    \centering

    \begin{subfigure}[t]{0.49\textwidth}
        \centering
        \includegraphics[width=\linewidth]{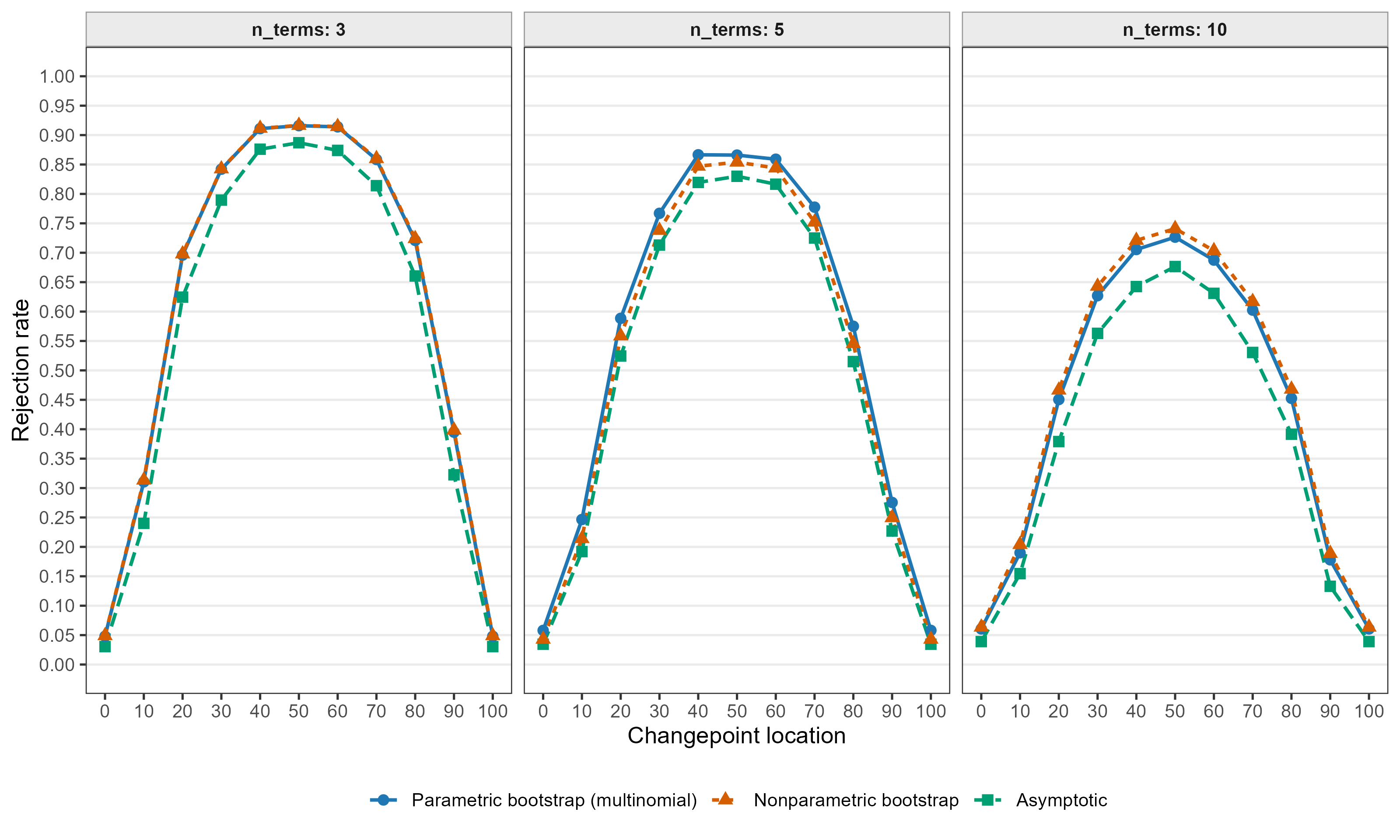}
        \caption{ Proposed test statistic, constant document intensity.}
    \end{subfigure}
    \hfill
    \begin{subfigure}[t]{0.49\textwidth}
        \centering
        \includegraphics[width=\linewidth]{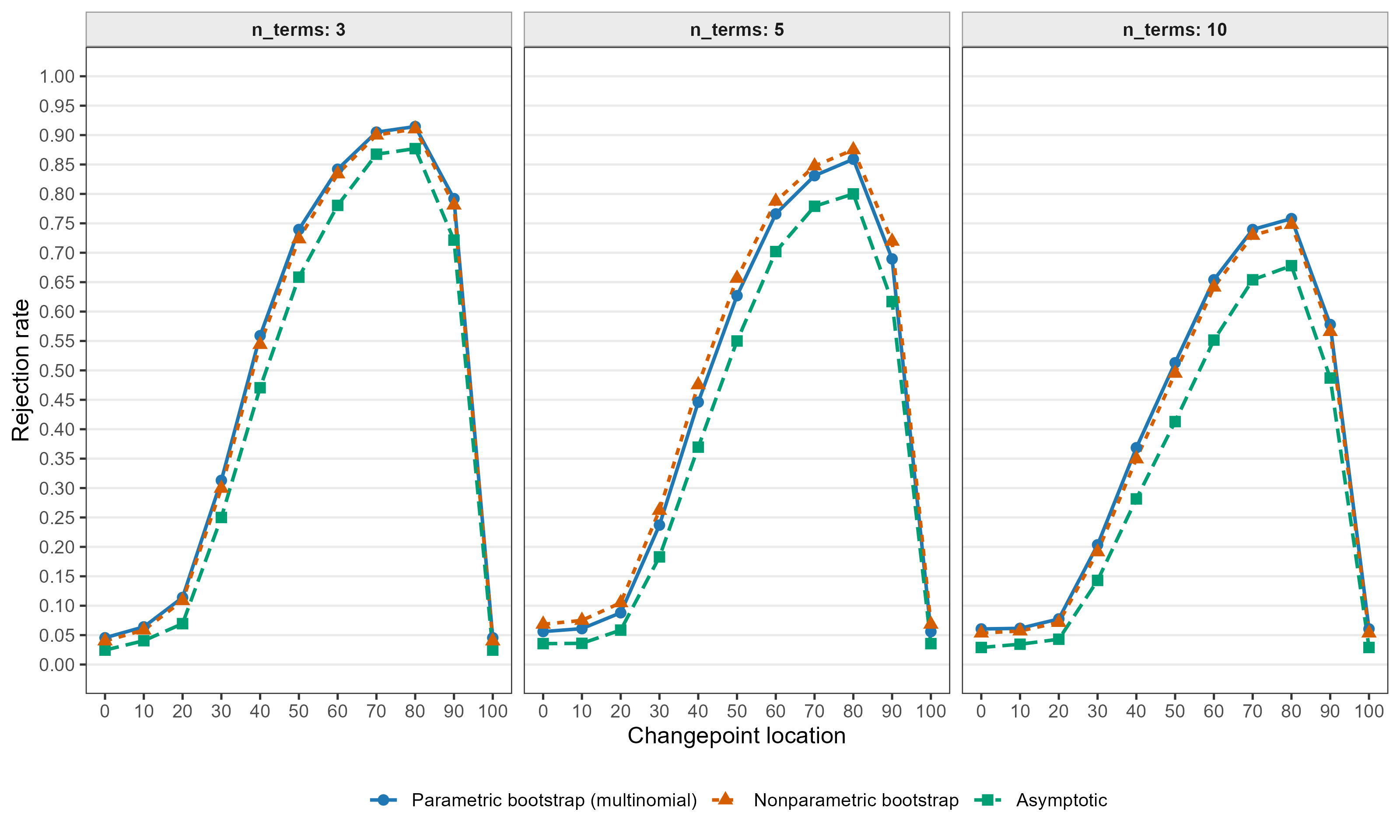}
        \caption{ Proposed test statistic, exponentially growing document intensity.}
    \end{subfigure}
\caption{Empirical rejection rates under multinomial data generation for the proposed test statistic under different calibration methods, with average monthly document intensity being equal 64. }
    \label{fig:multinomial_power_main_lambda64}
\end{figure}


\begin{figure}
    \begin{subfigure}[t]{0.49\textwidth}
        \centering
        \includegraphics[width=\linewidth]
        {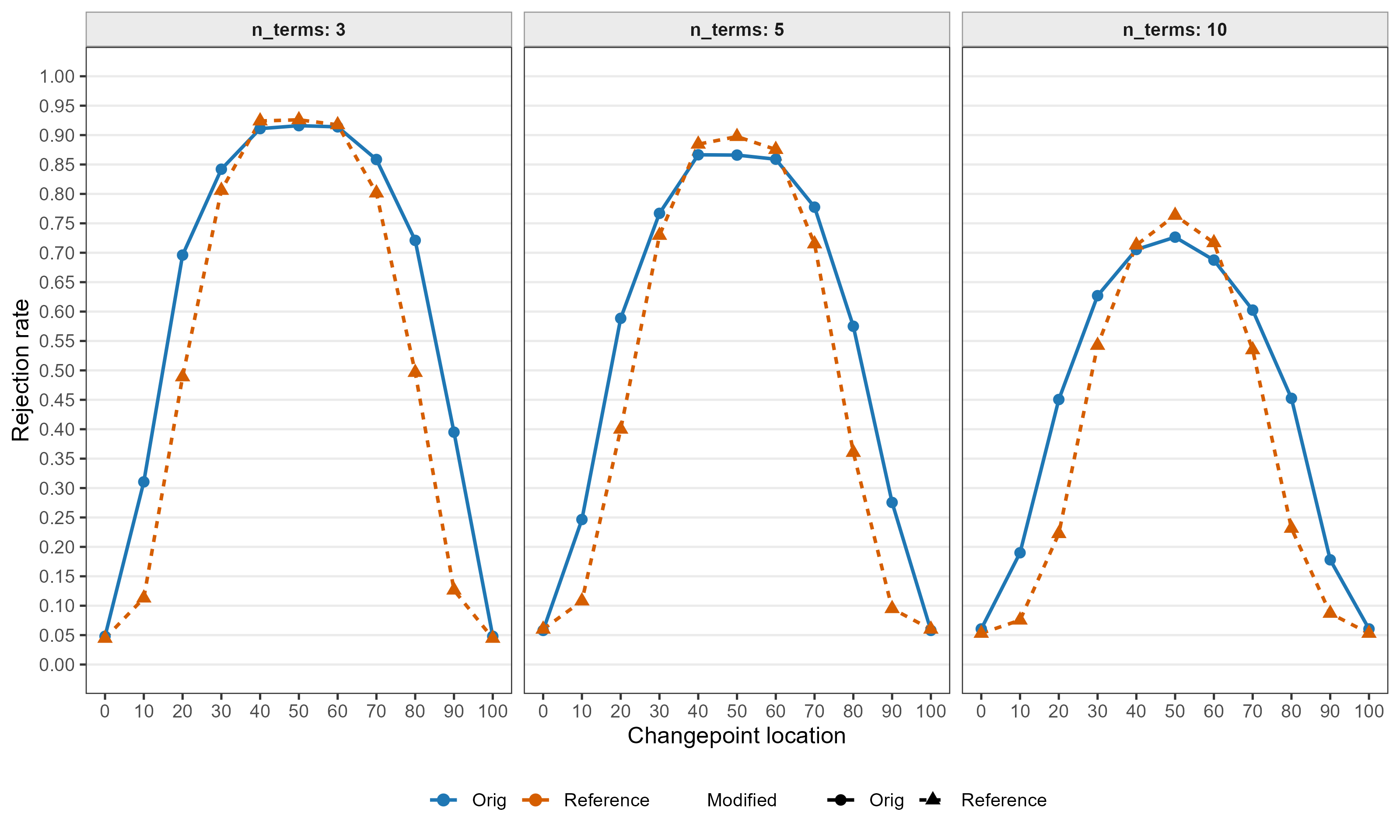}
        \caption{Parametric multinomial bootstrap, constant document intensity.}
    \end{subfigure}
    \hfill
    \begin{subfigure}[t]{0.49\textwidth}
        \centering
        \includegraphics[width=\linewidth]
        {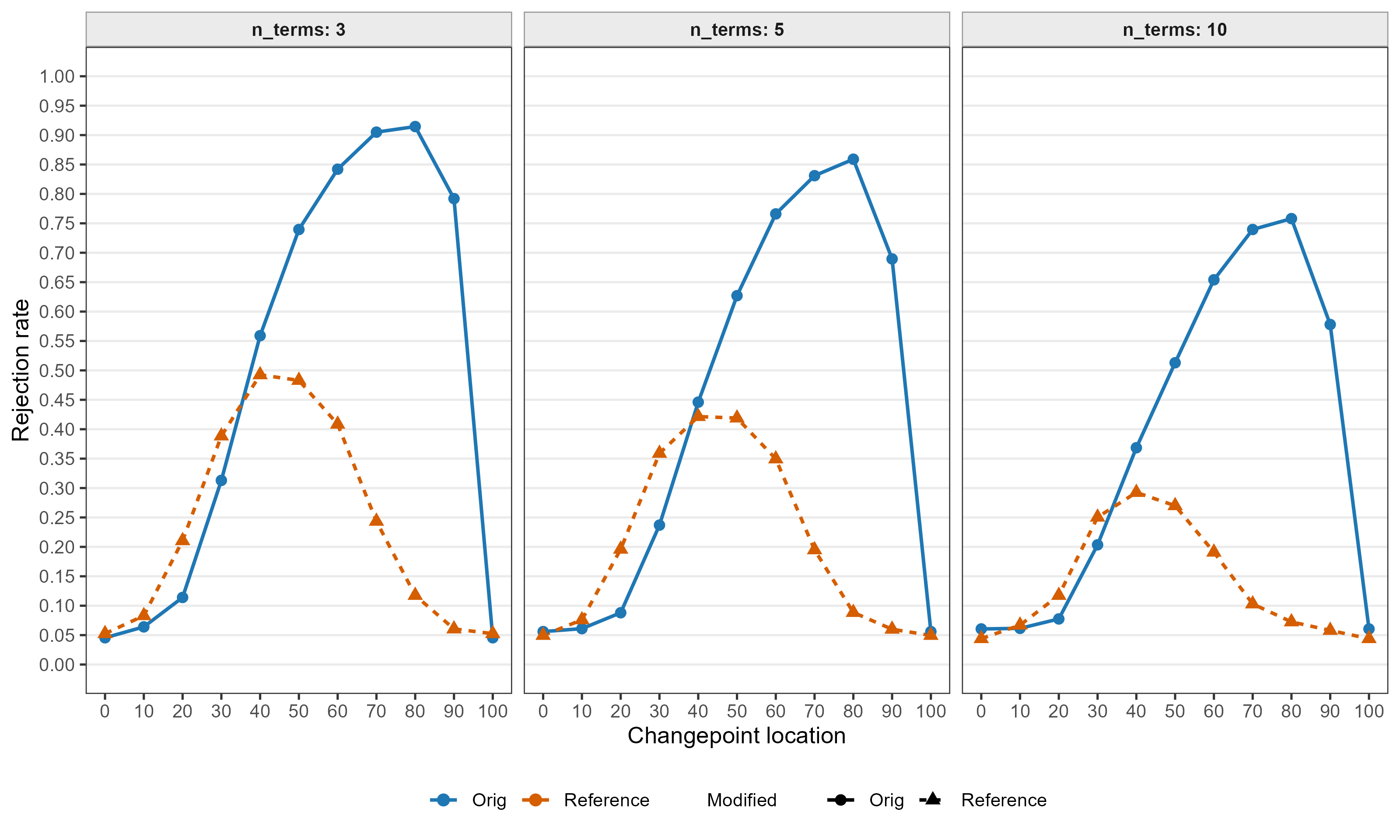}
        \caption{Parametric multinomial bootstrap, exponentially growing document intensity.}
    \end{subfigure}
    
    \vspace{0.5em}
    
    \begin{subfigure}[t]{0.49\textwidth}
        \centering
        \includegraphics[width=\linewidth]
        {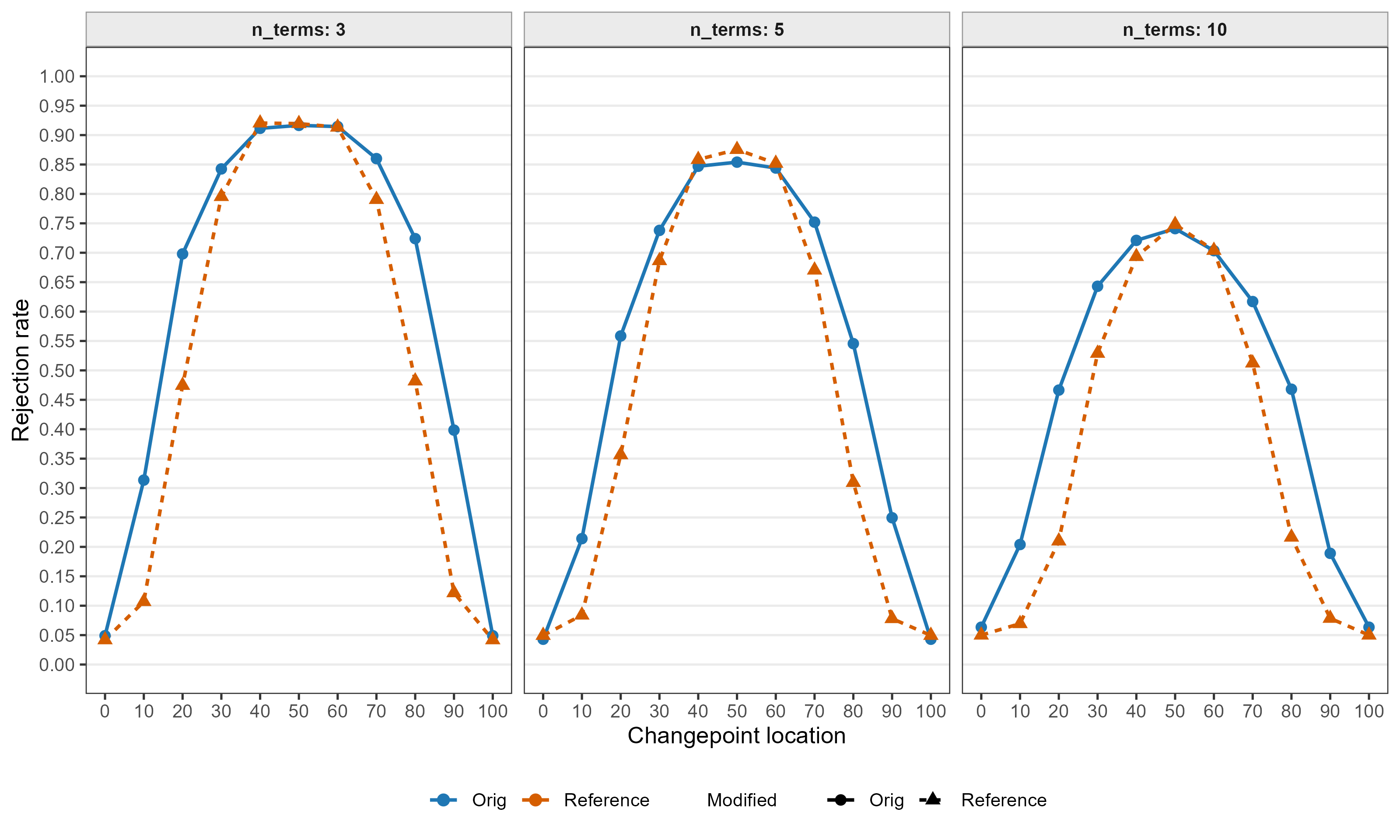}
        \caption{Nonparametric bootstrap, constant document intensity.}
    \end{subfigure}
    \hfill
    \begin{subfigure}[t]{0.49\textwidth}
        \centering
        \includegraphics[width=\linewidth]
        {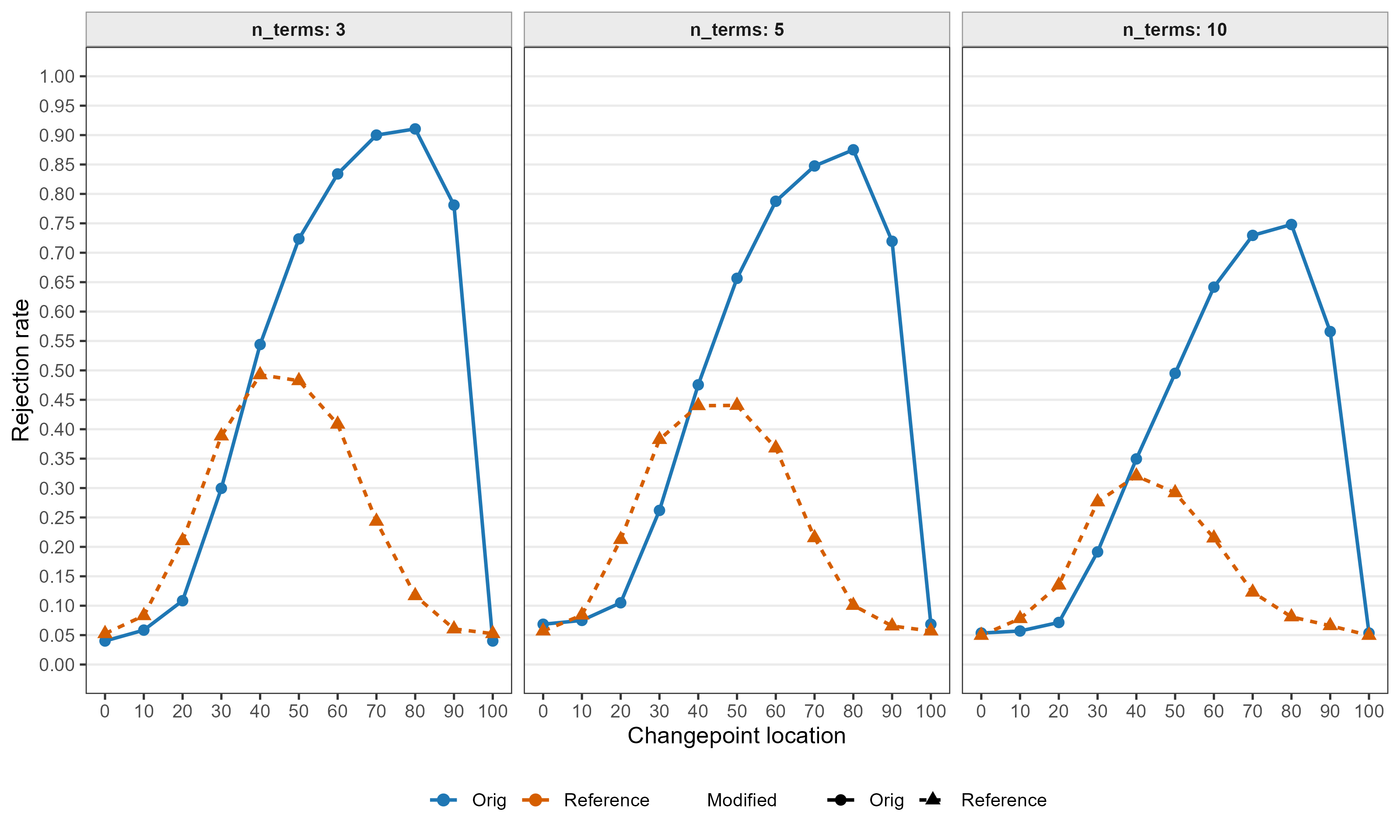}
        \caption{Nonparametric bootstrap, exponentially growing document intensity.}
    \end{subfigure}

    \caption{Empirical rejection rates under multinomial data generation - comparison to the competitor, with average monthly document intensity being equal 64.
    Panels (a)--(b) compare the different test statistics under multinomial bootstrap calibration,
    while panels (c)--(d) compare them under the
    using the nonparametric bootstrap. }
    \label{fig:multinomial_power_comp_lambda64}
\end{figure}

\begin{figure}[htbp]
    \centering

    \begin{subfigure}[t]{0.49\textwidth}
        \centering
        \includegraphics[width=\linewidth]
        {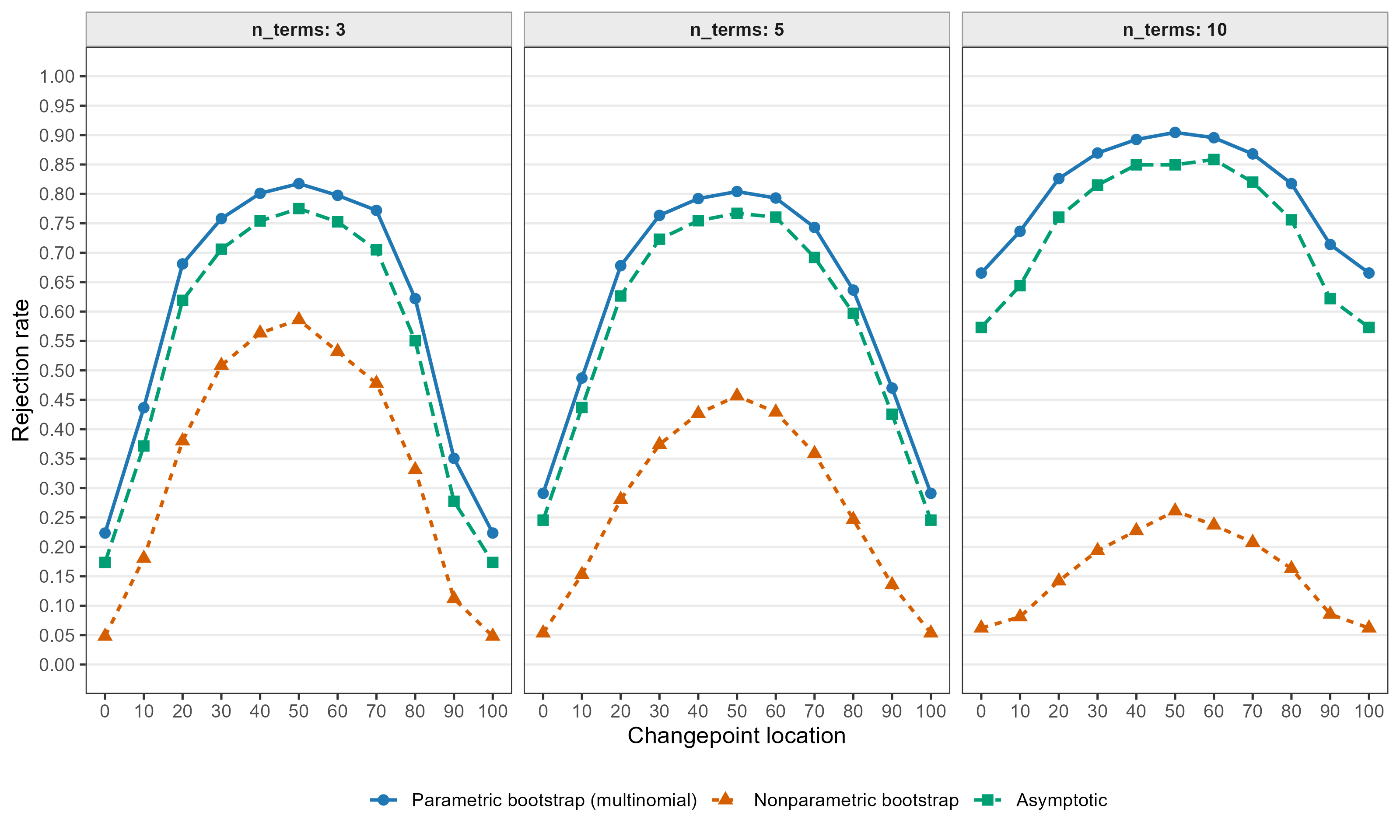}
        \caption{ Proposed test statistic, constant document intensity.}
    \end{subfigure}
    \hfill
    \begin{subfigure}[t]{0.49\textwidth}
        \centering
        \includegraphics[width=\linewidth]
        {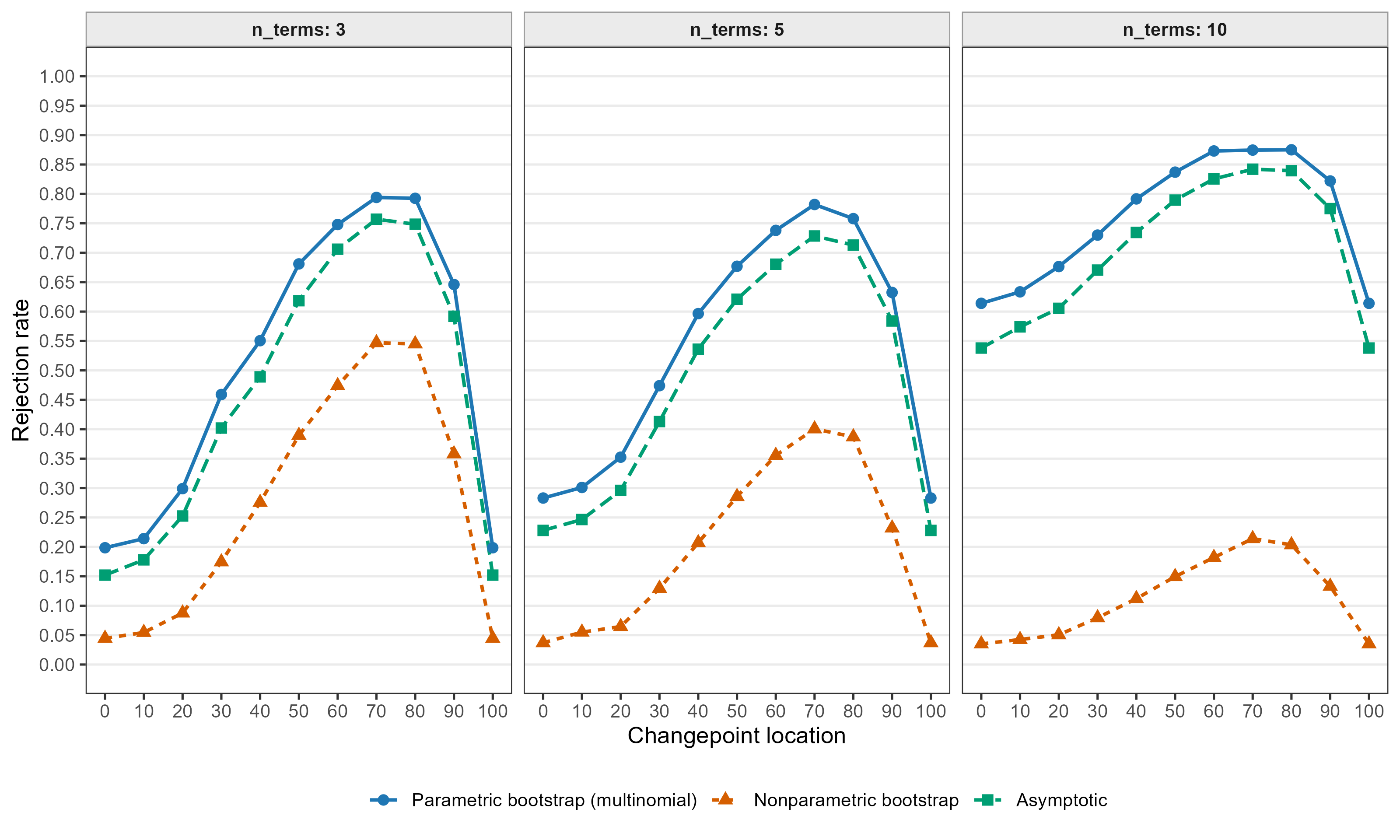}
        \caption{ Proposed test statistic, exponentially growing document intensity.}
    \end{subfigure}

    \begin{subfigure}[t]{0.49\textwidth}
        \centering
        \includegraphics[width=\linewidth]
{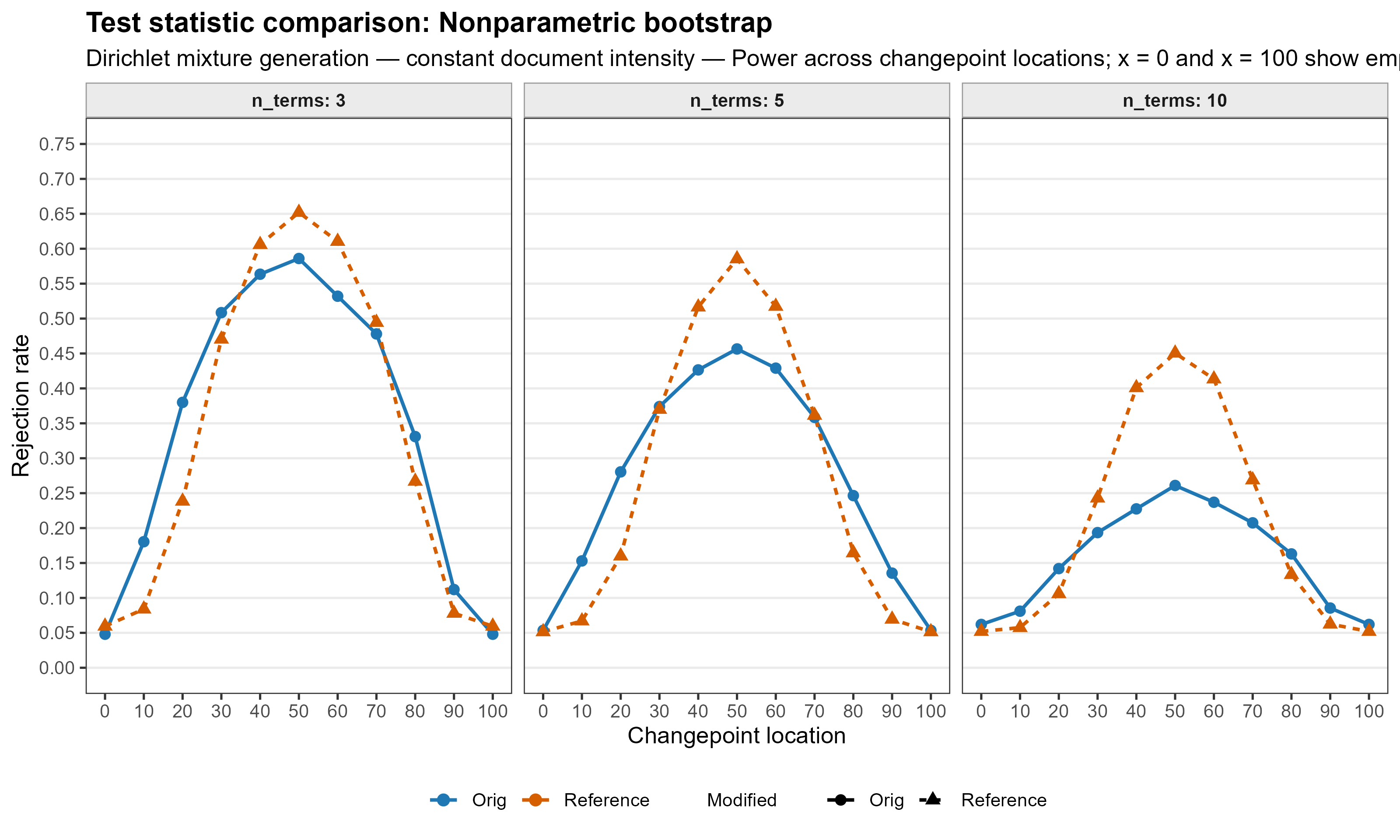}
        \caption{Constant document intensity.}
    \end{subfigure}
    \hfill
    \begin{subfigure}[t]{0.49\textwidth}
        \centering
        \includegraphics[width=\linewidth]
        {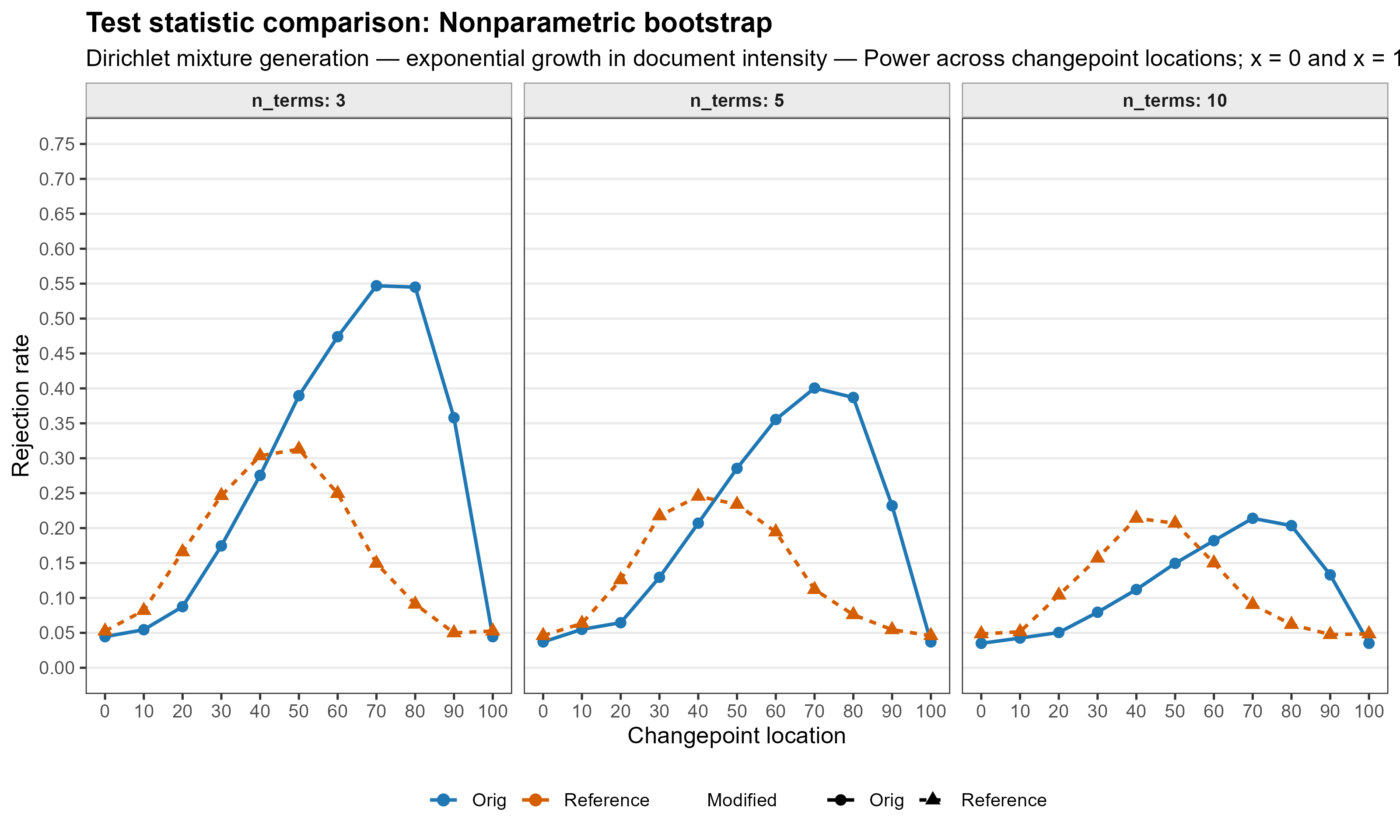}
        \caption{Exponentially growing document intensity.}
    \end{subfigure}
    
    \caption{Empirical rejection rates under Dirichlet mixture data generation - Average monthly document intensity being equal 16.
    Panels (a)--(b) compare calibration methods for the original test statistic,
    while panels (c)--(d) compare the proposed test statistic with the competitor using the nonparametric bootstrap.}
    \label{fig:dirichlet_power_main_lambda16}
\end{figure}

\begin{figure}[htbp]
    \centering

    \begin{subfigure}[t]{0.49\textwidth}
        \centering
        \includegraphics[width=\linewidth]
        {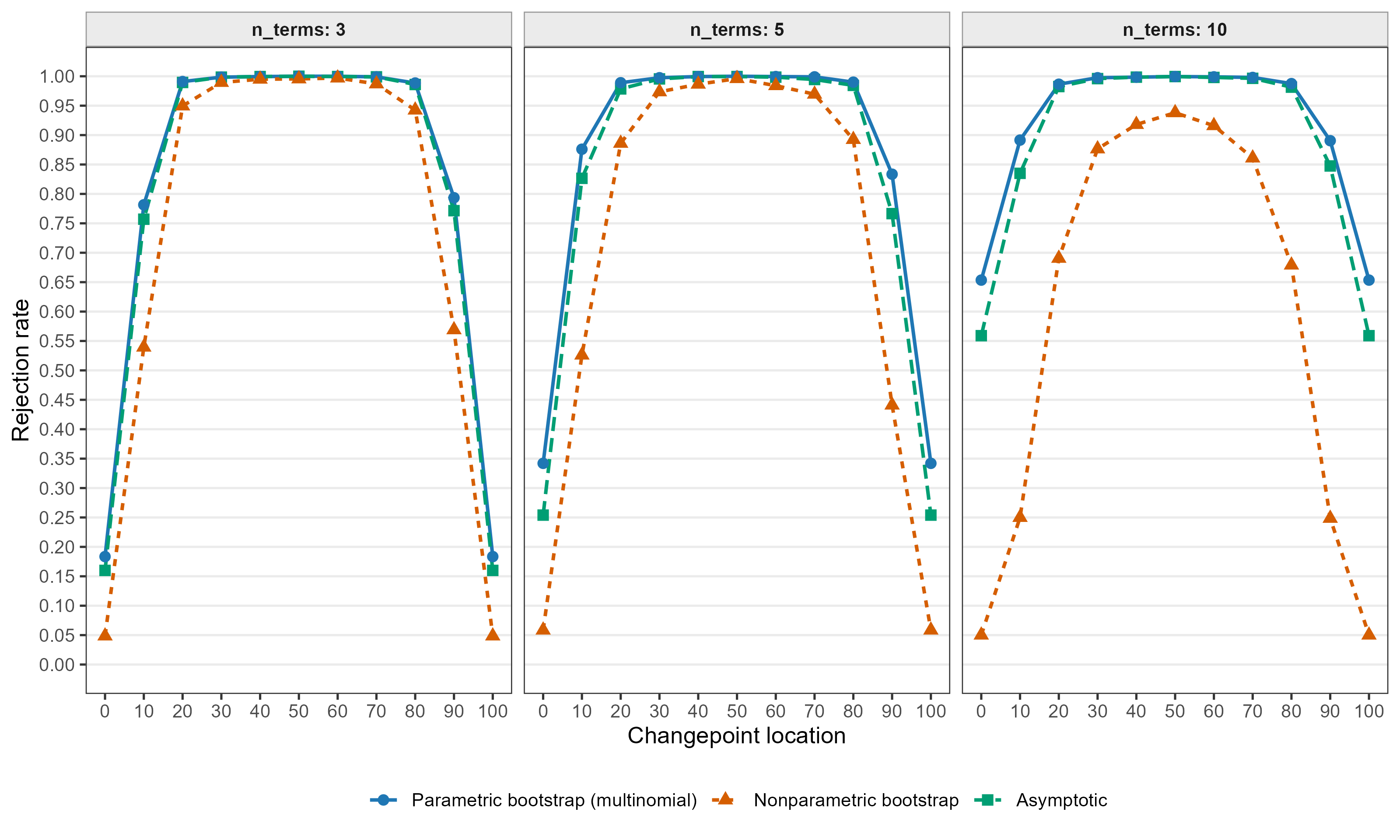}
        \caption{ Proposed test statistic, constant document intensity.}
    \end{subfigure}
    \hfill
    \begin{subfigure}[t]{0.49\textwidth}
        \centering
        \includegraphics[width=\linewidth]
        {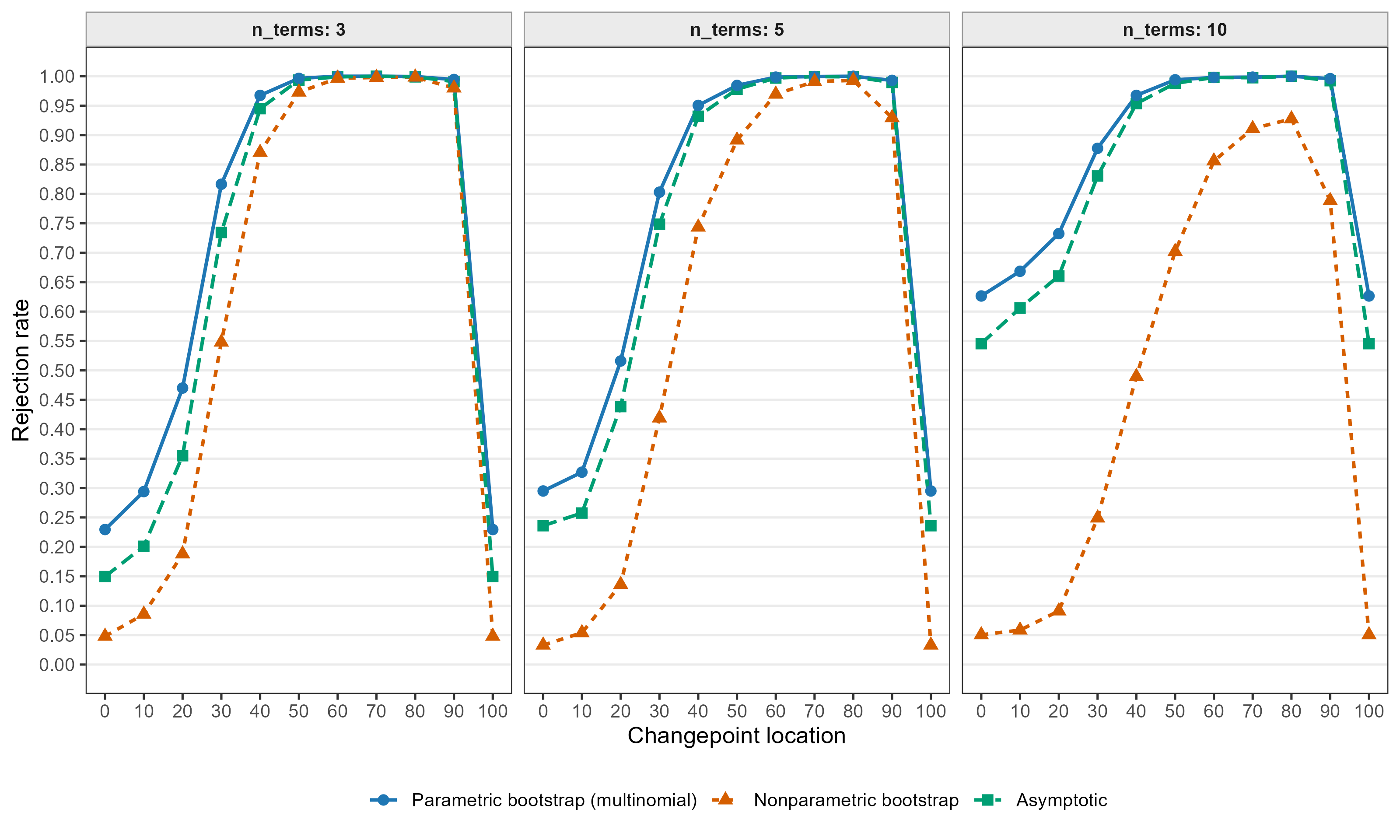}
        \caption{ Proposed test statistic, exponentially growing document intensity.}
    \end{subfigure}

    \vspace{0.5em}

    \begin{subfigure}[t]{0.49\textwidth}
        \centering
        \includegraphics[width=\linewidth]
{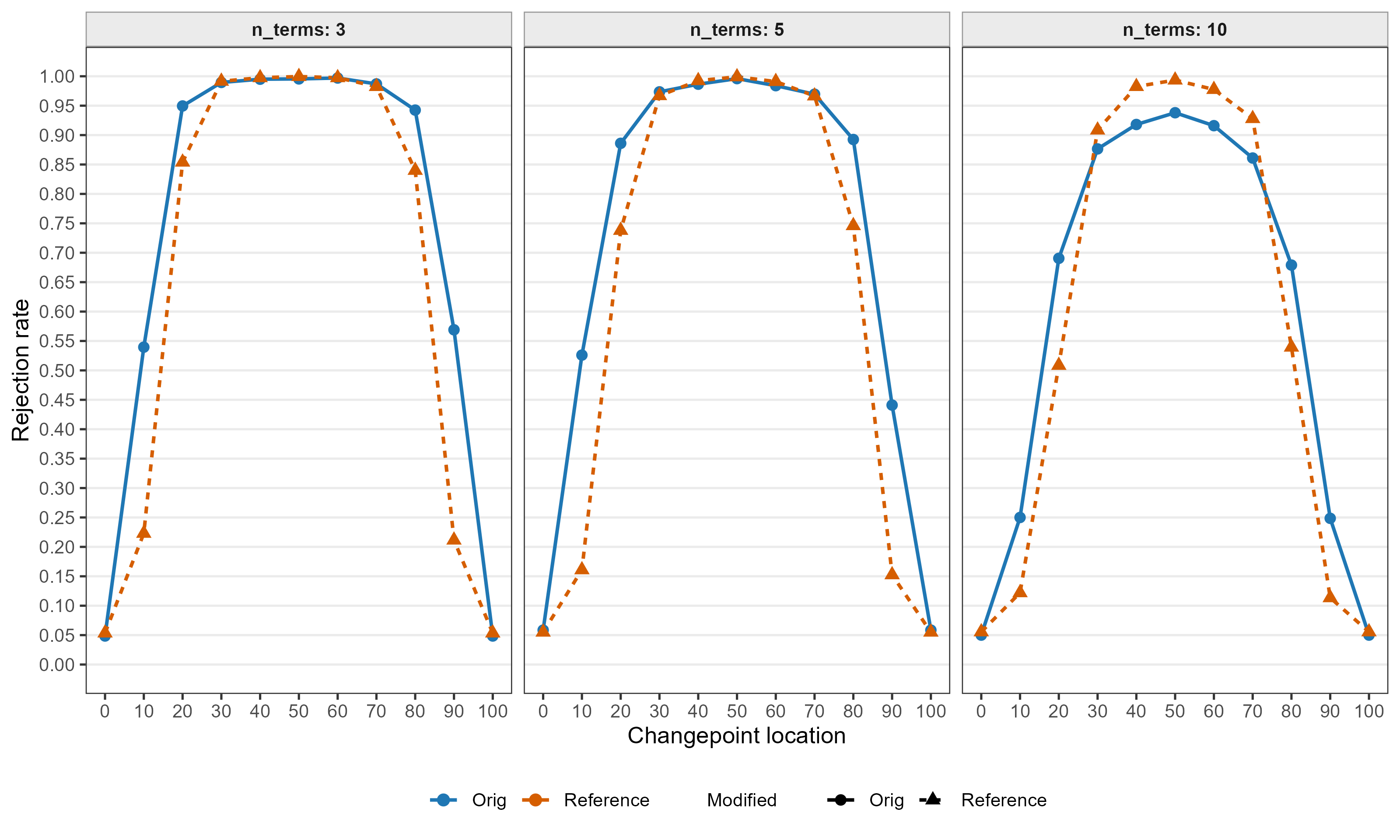}
        \caption{Constant document intensity.}
    \end{subfigure}
    \hfill
    \begin{subfigure}[t]{0.49\textwidth}
        \centering
        \includegraphics[width=\linewidth]
        {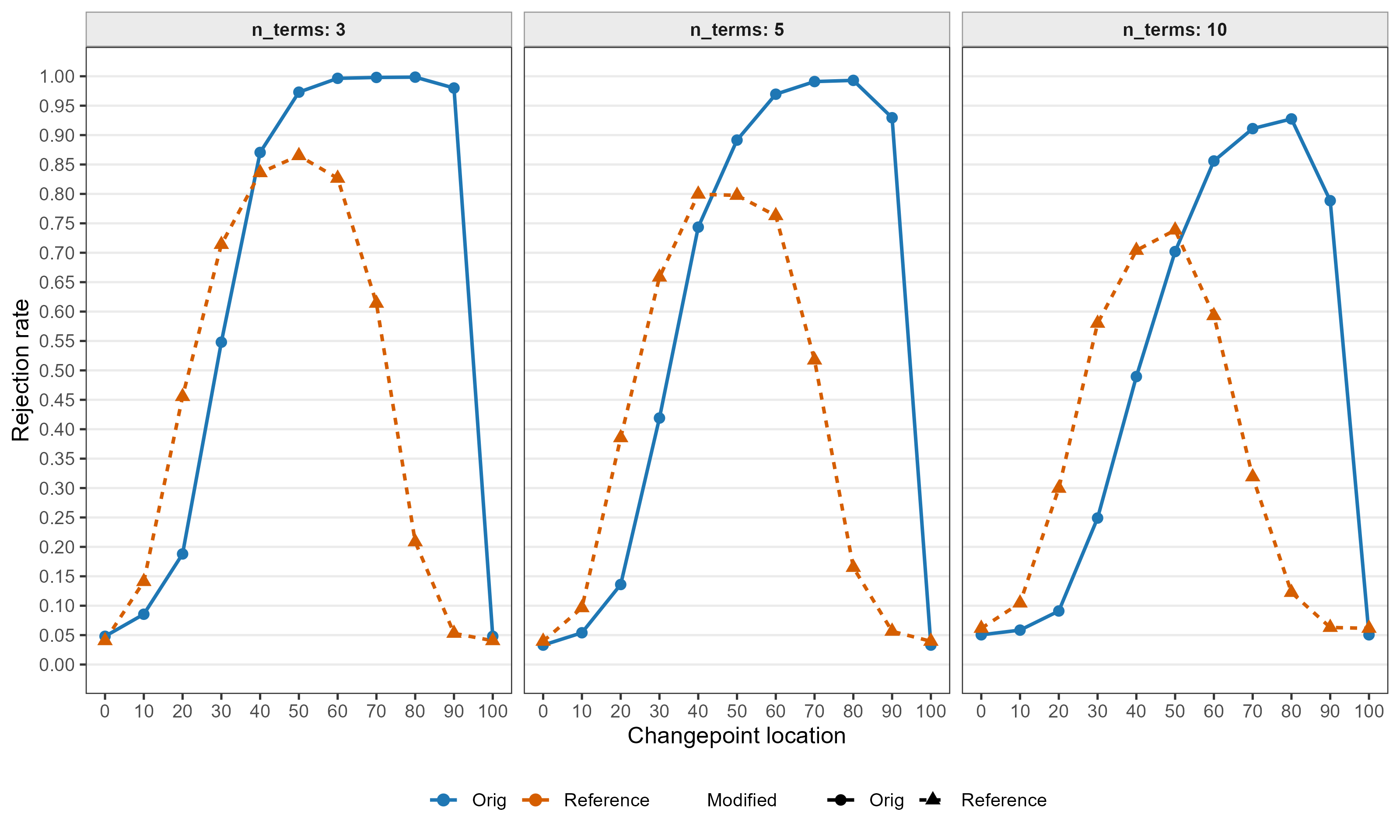}
        \caption{Exponentially growing document intensity.}
    \end{subfigure}
    
    \caption{Empirical rejection rates under Dirichlet mixture data generation - Average monthly document intensity being equal 64.
    Panels (a)--(b) compare calibration methods for the original test statistic,
    while panels (c)--(d) compare the proposed test statistic with the competitor using the nonparametric bootstrap.}
    \label{fig:dirichlet_power_comp_lambda64}
\end{figure}

\end{appendices}
\end{document}